\documentclass[11pt]{article}
\usepackage{verbatim}
\usepackage{amsmath,amssymb,amsthm,fullpage}
\usepackage{epsfig}

\newcommand{\GII} {\ensuremath{G_{II}}}
\newcommand{\GIII} {\ensuremath{G_{III}}}

\newcommand{\hG} {\ensuremath{G_{III}}}
\newcommand{\hf} {\ensuremath{f_3}}
\newcommand{\ZZ}{\mathbb{Z}}
\newcommand{\la}{\langle}
\newcommand{\ra}{\rangle}
\newcommand{\defd}{\ensuremath{=}}

\newcommand{\vval}{\ensuremath{\nu}}
\newcommand{\pval}{\vval}
\newcommand{\tval}{\ensuremath{\tilde{v}}}
\newcommand{\hide}[1]{}
\newcommand{\gvd}{``good'' vine decomposition}
\newcommand{\gvs}{vine selection}

\newcommand{\dto}{\ensuremath{\text{\textsc{Dto}}}}
\newcommand{\rto}{\ensuremath{\text{\textsc{Rto}}}}
\newcommand{\drand}{\rto}
\newcommand{\ddet}{\dto}
\newcommand{\far}{\ensuremath{\text{\textsc{Far}}}}

\newcommand{\lru}{\ensuremath{\text{\textsc{Lru}}}}

\newcommand{\opt}{\ensuremath{\text{\sc Opt}}}
\newcommand{\bel}{\opt}
\newcommand{\on}{\ensuremath{\text{\textsc{On}}}}
\newcommand{\fifo}{\ensuremath{\text{\textsc{Fifo}}}}

\newcommand{\maxfar}{\ensuremath{\text{\textsc{Maxfar}}}}

\newcommand{\rmark}{\ensuremath{\text{\textsc{Rmark}}}}

\newcommand{\e}{\ensuremath{\varepsilon}}
\newcommand{\ie}{\emph{i.e.}}
\newcommand{\defeq}{=} 
\newcommand{\compr}{r}
\newcommand{\obl}{\mathsf{obl}}

\newcommand{\lemlab}[1]{\label{lemma:#1}}

\newcommand{\proplab}[1]{\label{prop:#1}}

\newcommand{\theolab}[1]{\label{theo:#1}}

\newcommand{\seclab}[1]{\label{sec:#1}}
\newcommand{\deflab}[1]{\label{def:#1}}

\def\lemref#1{Lemma~\ref{lemma:#1}}

\def\propref#1{Proposition~\ref{prop:#1}}

\def\theoref#1{Theorem~\ref{theo:#1}}

\def\secref#1{Section~\ref{sec:#1}}

\def\defref#1{Definition~\ref{def:#1}}

\def\etal{\emph{et al.}}

\def\C{{\mathcal{C}}}
\def\D{\mathcal{D}}

\def\H{{\mathcal{H}}}

\def\PP{\mathcal{P}}
\def\Q{\mathcal{Q}}

\def\T{\mathcal{T}}

\def\V{\mathcal{V}}

\def\O{\mathcal{O}}

\newcommand{\epspic}[4]{
         \begin{figure}[t]
           \begin{centering}
            \framebox{\includegraphics{fig/#1.eps}}
           \end{centering}
           \caption{#3}
           \label{#4}
         \end{figure}
         }

\theoremstyle{plain}
\newtheorem{theorem}{Theorem}
\newtheorem{lemma}{Lemma}[section]

\newtheorem{proposition}[lemma]{Proposition}

\theoremstyle{definition}
\newtheorem{definition}[lemma]{Definition}

\newtheorem{example}[lemma]{Example}

\theoremstyle{plain}

\begin{document}

\title{Truly Online Paging with Locality of Reference}
\author{
Amos Fiat and Manor Mendel\thanks{Work done while the author was a
Ph.D. student in Tel-Aviv University. Current affiliation: The Open University of Israel.}\\
Department of Computer Science, Tel-Aviv University,\\
E-mail: {\tt fiat@tau.ac.il}, {\tt  mendelma@gmail.com}}
\date{}

\maketitle

\begin{abstract}

The competitive analysis fails to model locality of reference in
the online paging problem. To deal with it, Borodin {\etal}
introduced the access graph model, which attempts to capture the
locality of reference. However, the access graph model has a
number of troubling aspects. The access graph has to be known in
advance to the paging algorithm and the memory required to
represent the access graph itself may be very large.

In this paper we present truly online strongly competitive paging algorithms
in the access graph model that do not have any prior information on the
access sequence. We present both deterministic and randomized algorithms.
The algorithms need only $O(k \log n)$ bits of memory, where $k$ is the
number of page slots available and $n$ is the size of the virtual address
space. {\sl I.e.}, asymptotically no more memory than needed to store the
virtual address translation table.

We also observe that our algorithms adapt themselves to temporal changes in
the locality of reference. We model temporal changes in the locality of
reference by extending the access graph model to the so called
\emph{extended access graph model}, in which many vertices of the graph can
correspond to the same virtual page. We define a measure for the rate of
change in the locality of reference in $G$ denoted by $\Delta(G)$. We then
show our algorithms remain strongly competitive as long as $\Delta(G) \geq
(1+\e)k$, and no truly online algorithm can be strongly competitive on a
class of extended access graphs that includes all graphs $G$ with
$\Delta(G)\geq k- o(k)$.
\end{abstract}
\newpage

\tableofcontents

\newpage

\section{Introduction}

\subsection{The Paging Problem and Competitive Analysis}

The paging problem is a simplification of an optimization problem that appears
in computer systems with {virtual memory}. In the paging problem, memory is
partitioned into two: Small and fast memory, called \emph{real memory} vs.
large and slow memory, called {\em virtual memory}. The memory space is divided
into equal sized regions, called {\em pages}: $k$ real memory pages, and $n$
virtual memory pages. Usually $n$ is much larger than $k$.

Programs address the virtual memory, and the address translation mechanism
translates it to a real memory address. Requests for virtual pages that are
already in the real memory are called \emph{page hits}. Whenever a requested
virtual page is not in the real memory, a \emph{page fault} occurs and the
requested page is brought into the real memory. \emph{A page eviction strategy}
decides what page is to be evicted from the real memory in order to make room
for the requested page. The goal of the strategy is to minimize the number of
page faults, and the decisions should be made {\em online\/}, {\ie}, without
knowing the future requests. Such a strategy is called a {\em paging
algorithm}.

If the paging algorithm has the entire request sequence in advance, \ie, it is
not an online algorithm, a simple optimal solution due to Belady
\cite{Belady66} is as follows: Evict the page whose next use is furthest in the
future. This strategy is called {\bel}.

In a seminal paper, Sleator and Tarjan~\cite{SleTar85a} suggest using {\em
competitive analysis\/} to measure the performance of online algorithms. Let
$A(k,\sigma)$ denote the number of page faults a paging algorithm $A$ incurs on
the sequence $\sigma$ using a real memory with $k$ page slots, and starting
with no pages in real memory. If $A$ is a randomized algorithm then
$A(k,\sigma)$ is a random variable.

Competitive analysis compares the cost of a given online algorithm to the
optimal offline algorithm. In what follows, we describe the use of the
competitive measure in the context of paging for randomized paging
algorithms. We use the notion of the {\sl oblivious adversary}
\cite{BBKTW94} where the adversary knows the paging algorithm but not the
random coin tosses of the paging algorithm.

A randomized online algorithm {\on} is called \emph{strictly $r$--competitive}
if $E[\on(k,\sigma)]\leq r \cdot \bel (k,\sigma)$ for every request sequence
$\sigma$. The infimum of $r$ for which {\on} is $r$--competitive is called the
strict competitive ratio of {\on} and is denoted by $r_{\on}(k)$. {\on} is
called \emph{asymptotically $r$-competitive}, if there exists a constant $C\geq
0$, such that on any request sequence $\sigma$, $E[\on(k,\sigma)]\leq r \cdot
\bel (k,\sigma)+C$. The infimum of $r$ for which {\on} is asymptotically
$r$--competitive is called the asymptotic competitive ratio of {\on} and is
denoted by $r^\infty_{\on}(k)$. Obviously, $r^\infty_{\on}(k)\leq r _{\on}(k)$.

As shown in~\cite{SleTar85a}, the best deterministic strict competitive ratio
and the best deterministic asymptotic competitive ratio for paging with $k$
page slots are both equal to $k$. Fiat {\etal}~\cite{FKLMSY91} proved that the
asymptotic competitive ratio for randomized paging algorithms is $\Omega(\ln
k)$ and the strict competitive ratio for randomized paging algorithms is
$\O(\ln k)$.

\subsection{Locality of Reference} \seclab{locality}

Competitive analysis of paging algorithms does not model reality well. It
fails to distinguish between algorithms that perform very differently in
practice. For example, both ``Least Recently Used" (\lru) algorithm and
``First In First Out" (\fifo) algorithm have optimal deterministic
competitive ratio of $k$, but in practice {\lru} out-performs {\fifo}.
Furthermore, the ``observed competitive ratio" of {\lru} is usually only a
constant, {\ie}, on typical request sequences its performance is worse than
{\bel} by a constant  ($\approx 4$) multiplicative factor \cite{FR97}.

A partial explanation for these phenomena is that programs exhibit {\em
locality of reference\/}. Informally, locality of reference means that pages
requested in the near past are likely to be requested in the near future, and
at any moment there is usually a small set of pages likely to be requested. The
standard competitive analysis does not consider locality of reference, as it
treats all possible request sequences the same. Thus, the competitive ratio is
likely to be unrealistically  high  for algorithms that better exploit the
locality of reference in the request sequence.

Motivated by this observation, Borodin, Irani, Raghavan and
Schieber~\cite{BIRS95} suggest incorporating locality of reference into the
competitive analysis. In their model, the set of possible request sequences is
limited to only those derived from walks on a fixed {\em access graph}.

An access graph $G=(V, E)$ for a program is a graph that has a vertex for
each page in the virtual memory. Locality of reference is imposed by the
adjacency relationships in the graph: A page $v$ can be requested
immediately after a page $u$ only if there is an edge between $u$ and $v$ in
the access graph. Hence, the possible request sequences are limited to those
correspond to paths in the access graph. \emph{Here we consider only
undirected access graphs}.

The competitive ratio of a paging algorithm is now dependent on the access
graph $G$. Let $\text{paths}(G)$ denote the set of finite length paths in $G$.
Then
 \begin{eqnarray}
\lefteqn{r_{\on}(G,k) =} && \nonumber \\ && \inf  \{r:\ \forall
\sigma\in\text{paths}(G),\
E[\on(k,\sigma)] \leq r \cdot \bel (k,\sigma)\} ,\nonumber\\
\lefteqn{r^\infty_{\on}(G,k) =} && \nonumber \\ && \inf  \{r:\ \exists C\geq
0\;\forall \sigma\in\text{paths}(G),\
 E[\on(k,\sigma)] \leq r \cdot \bel (k,\sigma) +C
\}. \nonumber
\end{eqnarray}

We define the following terminology and notation: \begin{itemize} \item The
deterministic competitive ratio of a paging problem, $$r(G,k)= \inf _A
r_{A}(G,k),$$ where $A$ ranges over the \emph{deterministic} online paging
algorithms.
\item The deterministic asymptotic competitive ratio of a paging problem, $$r^{\infty}(G,k)= \inf _A
r^{\infty}_{A}(G,k),$$ where $A$ ranges over the \emph{deterministic} online
paging algorithms.
\item The
randomized competitive ratio of a paging problem, $$r_{\obl}(G,k)= \inf _A
r_{A}(G,k),$$ where $A$ ranges over the \emph{randomized} online paging
algorithms. The subscript $\obl$ indicates the usage of the oblivious adversary model.
\item The
randomized asymptotic competitive ratio of a paging problem, $$r_{\obl}(G,k)=
\inf _A r_{A}^{\infty}(G,k),$$ where $A$ ranges over the \emph{randomized}
online paging algorithms.
\end{itemize}

We are interested in {\em uniform} online algorithms that are given the
access graph $G$ as their input (before receiving the request sequence)
 and work in poly$(|G|,i)$ time for the $i$th
request. We adapt the convention from~\cite{IKP96} and define a uniform
online paging algorithm $A$  to be \emph{very strongly competitive} if its
competitive ratio for any paging problem is bounded from above by a fixed
linear function of the \emph{asymptotic} competitive ratio of the paging
problem. \emph{I.e.},
there exist $b_1,b_2\geq 0$ such that  for every $k$ and $G$,%
 \begin{equation} \label{eq:strongly-competitive}
   r_{A}(G,k)\leq
  \begin{cases}
  b_1 r^\infty (G,k)+b_2 & \text{ if $A$ is deterministic}\\
b_1r^\infty_{\obl} (G,k) + b_2 & \text{ if $A$ is randomized}.
 \end{cases}
 \end{equation}

See Section~\ref{sec:very-strong} for discussion on the choice of this
definition.

\subsection{Truly Online Algorithms}

A problematic aspect of previous algorithms for the access graph model, such as
those in \cite{BIRS95,IKP96,FK95}, is the assumption that the access graph is
given in advance. This assumption has the following obvious drawbacks:
\begin{itemize}
\item It is not clear how the paging algorithm gets hold of the access graph. One possible solution suggested is that information be gathered on
the program access graph during the compile phase, but this argument is only
partially satisfactory.
\item The storage requirements just to represent the access graph are at least
as large as a constant fraction of the virtual memory size and may even be a
constant fraction of the virtual memory size squared!
\end{itemize}

In contrast, algorithms such as {\lru}, {\fifo} \cite{SleTar85a}
and {\rmark} \cite{FKLMSY91}, that are ``oblivious" to the
underline access graph, do not have those problems. We call such
algorithms \emph{truly online algorithms}. 
\begin{definition}
A uniform online paging algorithm {\on} is called \emph{truly online} if it
does not get the underlying access graph as an input, and only gets the page
request sequence (in an online fashion). More formally, Let $A$ be a uniform
paging algorithm. Denote by $A(G,k)$ the startgey of this algorithm tailored for
access graph $G$ and cache of size $k$.
$A$ is called truly online if for any two access graphs $G_1$ and $G_2$ on
the same vertex set, any $k\in \mathbb{N}$, and any request sequence
$\sigma$ compatible with both $G_1$ and $G_2$, $A(G_1,k)$, and $A(G_2,k)$ produce
the same distribution when applied to $\sigma$.
\end{definition}

Classic paging algorithms such as {\lru} and {\fifo} are truly online but
not strongly competitive, as demonstrated in \cite{BIRS95}. The existence of
truly online very strongly competitive algorithms is not at all obvious.
Nonetheless, in this paper we present two paging algorithms, a deterministic
algorithm and a randomized algorithm, with the following desirable
properties:

\begin{enumerate}
\item \emph{Both algorithms are truly online and very strongly competitive}.
This implies that knowing the access graph is not necessary for ``almost
optimal" online algorithms.

\item
Storage requirements are only $O(k \log n)$ bits, compared to a na\"{\i}ve
implementation that stores the whole access graph and needs $\Omega( n^2 \log
n)$ bits in the worst case. Using randomization, we can reduce the space
requirement even further to $O(k \log k )$ bits.

\item
Both algorithms can be implemented fairly efficiently to deal with page hits.
In fact, their hardware requirements for implementing page hits are comparable
to the complexity of implementing {\lru} in hardware.\footnote{Processing of
page faults is more complicated than the processing required by {\lru}.
Arguably, this is less important, since page faults are relatively infrequent,
and are accompanied with a large I/O overhead anyway, so implementing the page
fault logic in software is relatively insignificant.}

\item
Both algorithms are adaptive. If the page sequence exhibits different behavior
over time,  the algorithms adapt to these changes. Unfortunately, locality of
reference as captured in the access graph model is fixed, and therefore the
access graph model does not explain all properties of our algorithms. In the
next section (Section~\ref{sec:refined}) we consider a model for ``changing
locality of reference", which reveals the full strength of our algorithms.
\end{enumerate}

\subsection{Refined Locality of Reference} \label{sec:refined}

We seek a model that allows one to deal with changing behavioral patterns of
the access sequence over time. For example, a compiler may run in many stages,
with entirely different local behavior in the different stages. Because much of
the execution of software is performed in the operating system (I/O
processing), a common access graph would show that certain pages are accessed
from all over the address space, essentially losing much of the information
about locality of reference.

To deal with this we allow multiple appearances of virtual page labels in the
access graph. The
same page label may appear on many different vertices. The access
sequence is constrained to obey the locality conditions imposed by
the edge relations in the graph. \emph{I.e.}, every access
sequence is derived from a path in the graph. We call this model
the \emph{extended access graph model}.

For a given extended access graph $G$ we define the parameter $\Delta(G)$ to be
the shortest path in $G$ between two different vertices labeled with the same
page. Observe that $\Delta(G)$ is the minimum number of requests to different pages needed to separate requests for the same page that have a different set of neighbors.
Intuitively, $\Delta(G)$  indicates ``how quickly" locality of reference
changes.

As we shall see, our algorithms perform quite well with respect to
$\Delta(G)$. Specifically, we show that our algorithms are very strongly
competitive with respect to the family of all extended access graphs with
$\Delta(G)\geq (1+\e)k$. We also prove an almost matching impossibility
result: there exists a family of extended access graphs with $\Delta(G)\geq
k-o(k)$, such that no truly online algorithm can be very strongly competitive
on this family.

\subsection{Very Strong competitiveness vs. Strong competitiveness}
\label{sec:very-strong}

The definition of very strong competitiveness
(Eq.~\eqref{eq:strongly-competitive}) uses  the strict competitive ratio on the
left hand side, yet makes use of the  asymptotic competitive ratio on the right
hand side.

 A more commonly used measure in previous
literature \cite{IKP96,FK95} is the following weaker notion of strong
competitiveness. A truly online algorithm $A$ is called \emph{strongly
competitive} if there exist $b_1,b_2\geq 0$ such that for every $k$ and $G$,
 \[ r^\infty_{A}(G,k)\leq
  \begin{cases}
  b_1 r^\infty (G,k)+b_2 & \ $A$\text{ is deterministic}\\
b_1r^\infty_{\obl} (G,k) + b_2 & \ $A$ \text{ is randomized}.
 \end{cases}
 \]

In this section we clarify our choice. First we note that the notion of very
strong competitiveness implies strong competitiveness. We also note that the
proofs of strong competitiveness in \cite{IKP96,FK95} actually imply very
strong competitiveness.

We give upper bounds on the strict competitive ratio and lower bounds on the
asymptotic competitive ratio. If one gets an upper bound on the strict
competitive ratio --- one also has an upper bound on the asymptotic competitive
ratio. Likewise, a lower bound on the asymptotic competitive ratio implies a
lower bound on the strict competitive ratio. Thus, our results are the
strongest possible amongst the various variants.

When considering uniform (non truly online) algorithms, one should be careful
when using strong competitiveness. In this case, a uniform algorithm could have
computed an optimally asymptotically competitive online strategy (see
\cite{BIRS95}) by amortizing a long computation in terms of $|G|$ over a long
prefix of the request sequence, and using a large constant additive term $C$ to
cover the cost incurred while processing the prefix of the request sequence.

Here we avoid this problem by using the strict competitive ratio. Any algorithm
with a ``good" strict competitive ratio avoids the potential pitfall of simply
waiting sufficiently long so as to learn the page request distribution.

We next argue that truly online strongly competitive algorithms easily follow 
from existing uniform algorithms appearing in~\cite{BIRS95,IKP96,FK95}.
Let $A$ be one of the uniform
algorithms from~~\cite{BIRS95,IKP96,FK95}. 
Execute $A$ on ``the observed access graph" so far, {\ie}
the graph that contains all edges that have been used thus far in the prefix
of the request sequence. The resulting algorithm is clearly truly online.

To see that the resulting algorithm is also strongly competitive, observe
that the algorithms of \cite{BIRS95,IKP96,FK95} have the
marking property, and furthermore, the proof that they are $O(r)$ competitive
uses the argument that on any phase with $g$ new pages, they fault at most $O(rg)$
times. Hence, when analyzing their truly online counterparts, we observe that
in phases in which no new edges of the access graph are revealed, these
algorithms fault at most $$g \times (\mbox{the competitive ratio of the access
graph observed thus far}).$$ In phases during which new edges of the access graph are
revealed, these algorithms fault at most $k$ times (as any marking algorithm). 
As there are at most
$\binom{n}{2}$ phases in which new edges of the access graph can be revealed, we can
use $C=O(n^2 k)$ as the constant additive term in the definition of asymptotic
competitive ratio
--- thus showing that these algorithms are strongly competitive.

This type of solution has the following drawbacks: (i) The additive term can
be huge, as $n$ is typically much larger than $k$. (ii) It requires
$\Omega(n^2)$ memory. (iii) It does not extend to the extended access graph
model. Therefore, in the reminder of this paper we will only consider very
strong competitiveness.

\subsection{Related Work}

Borodin {\etal}~\cite{BIRS95} introduce the access graph model. They present
some basic facts about it and investigate popular algorithms like {\lru} and
{\fifo} in this context. In particular, they prove that the competitive
ratio of {\lru} is at most twice the competitive ratio of {\fifo} for the
same access graph. They also show that {\lru} performs badly on access graphs
with cycles of size $k+1$. Later Chrobak and Noga \cite{CN99} proved that
{\lru} is better than {\fifo} in this model, \emph{i.e.},
$r^\infty_{\text{\lru}}(G,k) \leq r^\infty_{\text{\fifo}}(G,k)$ for any
access graph $G$ and $k$.

Borodin {\etal}~\cite{BIRS95} also consider deterministic uniform paging
algorithms. They prove the existence of an optimal paging algorithm in
PSPACE($|G|$). They give a natural uniform paging algorithm, called {\far},
and prove that {\far} obtains a competitive ratio no worse than $O(\log k)$
times the asymptotic competitive ratio for the graph. This result is
improved in a paper by Irani, Karlin and Phillips \cite{IKP96} in which it is
shown that {\far} is very strongly competitive. The same paper also presents
a very strongly competitive algorithm for a sub-class of \emph{directed}
access graphs, called \emph{tree connected directed cycles}.

Fiat and Karlin \cite{FK95} present a strongly competitive randomized
algorithm, and a strongly competitive algorithm for multi pointer paging
(where the page requests come from more than one source). The latter gives an
alternative deterministic strongly competitive algorithm. The algorithms of
\cite{FK95}, deterministic and randomized, are the basis of this paper.

Karlin, Phillips and Raghavan \cite{KPR00} consider a paging problem where
the input to the paging algorithm is a Markov chain with states correspond to
pages, and probabilities $(p_{i j})_{i j}$ such that $p_{i j}$ is the
probability page $j$ is referenced just after page $i$. They show a paging
algorithm that is within a constant multiplicative factor of the optimal
online algorithm when the request sequences are generated from the Markov
chain. A simpler and better algorithm for Markov paging and generalizations
was given by Lund, Phillips and Reingold \cite{LPR99}.

Fiat and Rosen \cite{FR97} present an access graph based heuristic that is truly
online and makes use of a (weighted) dynamic access graph. In this sense we
emulate their concept. While the Fiat and Rosen  algorithm is experimentally
interesting in that it seems to beat {\lru}, it is certainly not strongly
competitive, and is known to have a competitive ratio of $\Theta(k \log k)$.

Much of the above work is summarized in \cite[chap. 3--5]{BEY98}.

\section{Preliminaries} \label{sec:prelim}

A crucial concept in this paper is the phase-partitioning of the
request sequence.
\begin{definition} [Phase partitioning \cite{FKLMSY91}]
The request sequence is partitioned into disjoint contiguous subsequences,
called phases, as follows. The first phase begins at the beginning of the
sequence. The $i$th phase begins immediately after the $i-1$th phase ends,
and it ends either at the end of the sequence, or just before the request for
$k+1$'th distinct page during the $i$th phase (whatever comes first). Note
that phase partitioning can be done in an online fashion.
\end{definition}

A \emph{new page} for the $i$th phase is a page which has been requested in
the $i$th phase, and either $i=1$ or the page was not requested in
the $(i-1)$-phase . The following lemma clarifies the importance of the phase
partitioning of request sequences.

\begin{lemma} \label{lem:opt-faults}
\cite{FKLMSY91} Given a request sequences $\sigma$ composed of $\ell$
phases, where the $i$th phase has  $g_i$  new pages. Then, $g_i\geq 1$ and
 \[  \frac{1}{2}\sum_{i=1}^\ell g_i \leq \bel(\sigma,k) \leq \sum_{i=1}^\ell
 g_i. \]
\end{lemma}

In order to prove that an online algorithm is strictly $r-$competitive, 
it is therefore sufficient
to show that in a phase with $g$ new pages, the online algorithm faults at most
$\frac{rg}{2}$ times.

A page that has been already requested during the current phase is called
\emph{marked}. Marks are erased at the end of the phase. An online algorithm
is said to have the \emph{marking property} if it never evicts a marked
page.  The only difference between different marking algorithms is the page
eviction strategy used for unmarked pages. Note that marking algorithms have
at most $k$ faults in a phase. All the algorithms we consider in this paper
are marking algorithms.

For marking algorithms, we use the term \emph{hole} to denote a page that was
requested during the previous phase, evicted during the current phase and has
not been requested yet during the current phase. A page is called
\emph{stale} if it was requested in the previous phase, and has not yet been
requested or evicted in the current phase.

Borodin {\etal} \cite{BIRS95} present the following lower bounds on the
asymptotic competitive ratio. Let $\ell(T)$ denote the number of leaves in a
tree $T$, and let $\T_i(G)$ denote the set of $i$-vertex sub-trees of the
graph $G$.
\begin{lemma}[\cite{BIRS95}] \lemlab{tree-lb}
For any access graph $G$ and $k$ page slots,
\begin{align*}
\compr^\infty(G,k) &\geq \max \{ \ell(T)-1 \, |\, T\in \T_{k+1}(G) \}, \\
\compr^\infty_{\obl}(G,k) &\geq \max \{ H_{\ell(T)-1} \, |\, T\in
\T_{k+1}(G) \},
\end{align*}
where $H_n=\sum_{i=1}^n i^{-1}$.
\end{lemma}

We obtain an estimate of the number of leaves that can be found in a sub-tree
of a given graph $G$ by the following proposition, (see~\cite{KW91} and references therein).

\begin{proposition} \proplab{leaves}
Let $G=(V, E)$ be a connected graph with $k <|V| \leq 2k$ and
$\ell$ vertices with degrees other than two. Then there exists a
sub-tree of $G$ on $k+1$ vertices with at least $\ell /30$ leaves.
\end{proposition}

Borodin {\etal}~\cite{BIRS95} present another lower bound using
the notion of {\em vine decomposition}.
\begin{definition}\cite{BIRS95} \deflab{vine-decomp}
A \emph{vine decomposition} $\V=(B,\PP)$ of a graph $G$ is a connected
sub-graph $B$ together with a set of paths $\PP=\{p_1,p_2,\ldots\}$ in $G$ 
such that 
(i) the endpoints of paths in $\PP$
are adjacent to vertices in $B$; 
(ii) The set of vertices appearing in paths in $\PP$ is disjoint to $B$.
(iii) The paths in $\PP$ are pairwise disjoint in terms of vertices.
$B$ is called the \emph{backbone} of
$\V$. For a path (vine) $p$ denote by $|p|$ the number of vertices in $p$
\emph{plus one}, \ie, the number of edges in $p$ including those connecting
them to $B$. Define the \emph{value} of vine decomposition $\V=(B,{\PP})$ to
be $\vval(\V) \defeq \sum_{p\in \PP} \log|p|$.
\end{definition}

\begin{lemma} \cite{BIRS95} \lemlab{vinedecomp-lb}
Denote by $\H_i(G)$ the set of vine-decompositions of $i$-vertex
subgraphs of $G$.  Then, \( \compr^\infty(G,k) \geq \max
\{\vval(\V) \, | \, \V \in \H_{k+1}(G) \}.\)
\end{lemma}

The following lower bound on the asymptotic competitive ratio is useful when
the access graph contains a ``large" cycle.

\begin{lemma}\cite{IKP96} \lemlab{bigcycle-lb}
If $(B,\PP)\in\H_{k+g}(G)$  and $g\geq 1$, then
\[ \compr^\infty(G,k) \geq \bigl \lfloor\max_{p\in\PP} \log (|p|-1) - \log g
\bigr \rfloor /2 . \]
\end{lemma}

An analogous lower bound for randomized algorithms, due to Fiat and Karlin
\cite{FK95}:
\begin{lemma}\cite{FK95} \lemlab{vinedcomp-rlb}
For any $(B,\PP) \in {\cal H}_{k+g}(G)$ with at least $2g$ vertices in
$\PP$, where $g\geq 1$,
\[ \compr^\infty_{\obl}(G,k)=\Omega \bigl(\log (  \sum _{p\in {\PP}}  |p|)
  - \log g\bigl) .\]
\end{lemma}

The following proposition is immediate.
\begin{proposition}
\label{lemma:subgraph} If $G$ is a sub-graph of $G'$ then
$r^\infty_{\obl}(G,k) \leq r^\infty_{\obl}(G',k)$ and
$r^\infty(G,k) \leq r^\infty(G',k)$.
\end{proposition}

\section{Randomized Algorithms} \seclab{rto}

Our algorithms are similar to Fiat and Karlin's algorithms
\cite{FK95}, but they do not have the access graph available in
advance. Instead, they make use of a spanning tree of the graph
resulting from the request sequence of the previous phase.

Let $P$ denote the pages requested in the previous phase. Let
$G_P=(P, E)$, where $E=\{u v | u, v\in P$ requested successively
in the previous phase, and $u\neq v\}$. Let $G_0=(V_0 \!\! = \!\!
P, E_0)$ be a spanning tree of $G_P$. Let $r_0$ denote the last
page requested in the previous phase and let $r_i$, $i\geq 1$,
denote the $i$th page request in the current phase. Define
$G_{i+1}=(V_{i+1},E_{i+1})$, where $V_{i+1}=V_i \cup \{r_{i+1}\}$,
and $E_{i+1}=E _i \cup \{ r_i r_{i+1} \} \setminus \{r_i r_i\}$.

Similar to Fiat and Karlin's algorithms, our algorithms are marking
algorithms having three sub-phases in a phase, with a different page eviction
strategy in each one of them. Let $\GII$ denote $G_i$ at the end of sub-phase
II, and $\GIII$ denote $G_i$ at the end of the phase.

In this section we present and analyze {\drand}, a truly online randomized
paging algorithm. In the first two sub-phases of {\drand}, a
vine-decomposition is constructed in $\GII$ such that the backbone of the
vine-decomposition consists of the marked vertices and the evicted vertices.
The third sub-phase evicts vertices randomly from the paths of the
vine-decomposition above.

\paragraph{Algorithm {\rto}$(k)$.} {\rto} is a marking algorithm that
partitions the phase into three consecutive sub-phases. In each
sub-phase it does as below. We emphasize that the graph referred
to in the following discussion is $G_0$, {\sl i.e.}, a spanning
tree of $G_P$.

\begin{description}
\item[Subphase I:] Denote by $C$ the set of vertices of degree not equal to 
two in $G_0$.
On a fault, evict a random unmarked unevicted (stale) page $v\in C$. If
there is no such page, and the phase is not over, proceed to sub-phase~II.

\item[Subphase II:] At the beginning of the sub-phase, all stale pages lie
on degree-2 vertices in $G_0$. Denote by $C'\subseteq C$, the set of holes
at the beginning of the subphase II. For each $v\in C'$, we maintain a
dynamic set $A_v$. At the beginning of the subphase, $A_v=\{v\}$. A vertex
$v\in C'$ is called ``alive" as long as $A_v$ contains only holes, and there
exists a stale vertex adjacent to $A_v$ in $G_0$.

On a fault choose $v$ such that
\begin{enumerate}
\item  $v\in C'$ and ``alive".
\item  $v$ minimizes $|A_u|$ amongst all candidate vertices $u$
        meeting condition 1 above.
\end{enumerate}
If no $v$ meets the criteria above, proceed to sub-phase~III.
Otherwise, evict a stale vertex $w$ adjacent to $A_v$, and set $A_v\leftarrow
A_v \cup\{w\}$.

\item[Subphase III:]
On a fault, evict a random unmarked page.
\end{description}

\subsection*{Competitive Analysis}

The analysis of {\rto} follows the analysis of the randomized algorithm from
\cite{FK95}.

Assume that $g$ new pages are requested during the phase, and let $f_i$
denote the expected number of pages evicted during sub-phase $i$ that will be
requested (later) during the phase. We note that the expected total number of
faults in the phase is at most $f_1+f_2+f_3+g$. We will show that $f_i=O(g
\cdot \compr^\infty_\obl(\GII,k))$. As $\GII$ is a sub-graph of the
underlying access graph $G$, it follows that {\drand} is very strongly
competitive on $G$.

\paragraph{Sub-phase I faults:} Let $S\subseteq C$ be the set of
vertices evicted in sub-phase~I. Suppose the adversary requested
the vertices in $S$, during the phase in the order
$s_1,s_2,\dots,s_m$.

\begin{proposition}
The probability that {\drand}  has a hole at $s_i$ at the time it is
requested, is at most $g/(|C|-i+1)$.
\end{proposition}
\begin{proof}
{\drand} has, at any point in time, at most $g$ holes. It is easy to prove by
induction on the number of requests since the beginning of the phase, that
after $s_{i-1}$ was requested and before $s_i$ will be requested, the holes
are evenly distributed among $C\setminus\{s_1,\dots,s_{i-1}\}$. For requests
during sub-phase I, this follows from the page eviction strategy of sub-phase
I. For requests after sub-phase I, $C\setminus \{s_1,\dots, s_i\}$ are all
holes, and $|C\setminus \{s_1,\dots, s_i\}|\leq g-1$. In either case, the
probability for a hole in $s_i$ is at most $g/(|C|-i+1)$.
\end{proof}

The expected number of evictions on $C$ is therefore no more than $g H_{|C|}$.
$G_1$ is a tree on $k+1$ vertices with $\Omega(|C|)$ leaves and by
\lemref{tree-lb}, $\compr^\infty_{\obl}(G_1,k) = \Omega (H_{|C|})$. Thus,
$f_1 = O(g \cdot \compr^\infty_{\obl} (G_1,k))$.

\paragraph{Sub-phase II faults:}
Note that $|C'|\leq g-1$. Let $a_v$ denote $|A_v|$ at the time immediately
before $v\in C'$ dies. Note that $f_2= \sum _{v\in C'} (a_v-1)$. Note also
that throughout sub-phase II: (i) $\sum_v |A_v| <g$ where the sum ranges
over the ``live" vertices in $C'$, and (ii) $\bigl||A_v| - |A_u|\bigr| \leq
1$ for any two ``live" vertices $u,v\in C'$. Denote by $v_i$ the $i$th vertex
that dies in $C'$. We conclude that $a_{v_i} \leq \left\lceil
g/(|C|-i+1)\right\rceil $. So $f_2\leq g+ g H_{|C|}=O(f_1)=O(g \cdot
\compr^\infty_{\obl} (G_1,k))$.

\paragraph{Sub-phase III faults:}
Denote by $\Pi$ the set of stale pages at the beginning of the
sub-phase. The vertices in $\Pi$ have degree 2 in $\GII$, since
any $u$ s.t. $\deg _{\GII}(u) \neq \deg _{G_0}(u)$ must have been
marked by now. Denote by $B$ the sub-graph of $\GII$ induced on
$(P \cup P') \setminus \Pi$, where $P$ is the set of vertices
requested in the previous phase, and $P'$ is the set of vertices
requested in sub-phases I and II.

\begin{proposition} \proplab{rto-B-connected}
$B$ is connected.
\end{proposition}
\begin{proof}
The vertices in $P'$ are on some path in $\GII$, and since $\Pi$ contains no
vertex from $P'$, we deduce that $P'$ is contained in some connected
component of $B$.

Assume, for the sake of contradiction, that $B$ has more than one connected
component. So, there must be another connected component $X$. Observe that $X$ must
intersect $C$, since $P \setminus \Pi \subset C$. 
For $v\in X\cap C$ let $A'_v$ be $A_v$ at the time $v$ dies.
Note that $X=\cup_{v\in X\cap C}A'_v$. As $X$ does not contain marked pages,
the only way the vertices in $X\cap C$ died is by not having any stale page
(page in $\Pi$) adjacent to $X$. A contradiction.
\end{proof}

Denote by $\overline{\Pi}$ the set of paths induced by $\Pi$ in $\GII$. The
endpoints of the paths in $\overline{\Pi}$ are adjacent (in $\GII$) to $B$.
Thus, from \propref{rto-B-connected}, $(B,\overline{\Pi})$ is a
vine-decomposition of a subgraph of $\GII$ with at most $k+g$ vertices. Denotes by
$L$ the number of vertices in $\Pi $. Denote by $g'$ the number of
holes on $\Pi$ at the end of the phase. Clearly, $g' \leq g$, and at most
$L-g'$ vertices from $\Pi$ can be requested during this sub-phase. As in
sub-phase I, the probability that the $i$th requested vertex in $\Pi $
($1\leq i \leq L- g'$) is a hole, is at most $g'/(L-i+1)$. Thus,
 \[ f_3\leq g'(H_L-H_{g'})\leq g'(\ln L - \ln g'+1) \leq g(\ln L - \ln g+1). \]
By \lemref{vinedcomp-rlb}, $\compr^\infty_{\obl}(\GII,k)
=\Omega(\log L -\log g)$. Therefore $f_3=O(g \cdot
\compr^\infty_{\obl} (\GII,k))$. We conclude,
\begin{theorem} \theolab{rto}
{\drand} is very strongly competitive on any underlying access graph.
\end{theorem}

\section{Deterministic Algorithms}

Next, we present {\ddet}, a deterministic truly online algorithm paging
algorithm. {\ddet} is similar to the deterministic algorithm from
\cite{FK95}, but instead of using a known access graph $G$, it makes use of
the dynamic tree $G_0$. As in {\drand}, the first two phases construct a
vine-decomposition in $\GII$ such that requested and evicted vertices form
the backbone of the vine-decomposition. Here, however, in Subphase III
{\ddet} attempts to evict pages laying in the middle of paths of unmarked
vertices.

As in {\drand}, all graph relations described herein relate to $G_0$. A path
$p$ in the graph satisfying certain property $P$ is called maximal (with
respect to containment) if there is no path $q$ the properly contains $p$ and
also satisfies $P$. A midpoint of a path is a  vertex or an edge that is
exactly in the middle of the path, \ie, at equal distance from both its
endpoints.

\begin{description}
\item[Subphase I:] Denote by $C$ the set of vertices of degree not equal two in $G_0$.
On a fault, evict an unmarked unevicted (stale) page $v\in C$. If there is
no such page, and the phase is not over, proceed to sub-phase~II.

\item[Subphase II:] At the beginning of the sub-phase, all stale pages lie
on degree-2 vertices in $G_0$. Denote by $C'\subseteq C$, the set of holes
at the beginning of the subphase II. For each $v\in C'$, we maintain a
dynamic set $A_v$. At the beginning of the subphase, $A_v=\{v\}$. A vertex
$v\in C'$ is called ``alive" as long as $A_v$ contains only holes, and there
exists a stale vertex adjacent to $A_v$.

On a fault choose a live $v\in C'$. If no $v$ meets this criteria, proceed
to the next sub-phase. Otherwise, evict a vertex vertex $w$ adjacent to
$A_v$, and set $A_v\leftarrow A_v \cup\{w\}$.

\item[Subphase III:]
On a fault, choose a maximal (w.r.t. containment) path $p$ of unmarked
vertices that contains a stale page, and evict a stale page in $p$ which is
closest to the midpoint of $p$.
\end{description}

\subsection*{Competitive Analysis}
Let $g$ be the number of new pages requested during the entire
phase, and let $f_i$ denotes the number of pages evicted during
sub-phase $i$. As in the case of {\drand}, the total number of
faults in the phase is at most $f_1+f_2+f_3+g$. We will show that
$f_i=O(g \cdot \compr^\infty(\GIII,k))$.

\paragraph{Sub-phases I \& II.}
Let $C\subseteq P$ be the set of vertices in $P$ with degree $\neq 2$ in
$G_0$. As in the analysis of {\drand}, denote by $a_v$ the size of $A_v$
immediately before $v\in C$ ``dies" . Since $f_1+f_2 = |C| + \sum_{v\in C'}
(a_v-1)$ and $a_v \leq g$, we conclude that $f_1+f_2 \leq g \cdot |C|$. From
\propref{leaves}, $G_1$ is a tree on $k+1$ vertices with $\Omega(|C|)$
leaves and by \lemref{tree-lb}, $\compr^\infty(G_1,k)= \Omega(|C|)$, and
therefore $f_1+f_2= O(g \cdot \compr^\infty(G_1,k))$.

\paragraph{Sub-phase III faults:}
Denote by $\Pi$ the set of stale pages at the beginning of the
sub-phase. The vertices in $\Pi$ have degree 2 in $\GII$, since
any $u$ s.t. $\deg _{\GII}(u) \neq \deg _{G_0}(u)$ must have been
marked by now. Denote by $B$ the sub-graph of $\GII$ induced on
$(P \cup P') \setminus \Pi$, where $P$ are the vertices requested
in the previous phase, and $P'$ are the vertices requested in
sub-phase I and II.
\begin{proposition}
$B$ is connected.
\end{proposition}
\begin{proof} Similar to the proof of \propref{rto-B-connected}.
\end{proof}
Denote by $\overline{\Pi}$ the set of paths induced by $\Pi$. The endpoints
of the paths in $\overline{\Pi}$ are adjacent to $B$, hence
$(B,\overline{\Pi})$ is a vine decomposition of $\GII$. Unlike the case in
\cite{FK95}, where the algorithm faults at most $g \log |p| $ times on every
path $p \in \overline{\Pi}$, {\ddet} might fault on \emph{every} vertex of
\emph{every} path $p\in \overline{\Pi}$. Nonetheless,  in
\secref{bounding-f3} we prove:

\begin{lemma} \lemlab{bounded-f3}
\label{LEMMA:BOUNDED-F3} \( f_3 = O(g \cdot \compr^\infty(\GIII,k)). \)
\end{lemma}

\noindent We conclude:
\begin{theorem} \theolab{dto}
{\ddet} is very strongly competitive on any underlying access graph.
\end{theorem}

\section{Proof of Lemma \ref{LEMMA:BOUNDED-F3}} \seclab{bounding-f3}

Our proof of \lemref{bounded-f3} is quite lengthy. To make the exposition
simpler it is partitioned as follows: \secref{informal} presents the
complications in  proving the lemma and gives some intuition.
\secref{preliminaries} introduces the notation used throughout the proof.
\secref{proof f_3} provides the proof, leaving out some combinatorial
lemmas. \secref{combinatorial} ends the exposition by providing the missing
proofs.

\subsection{Informal Exposition}
\seclab{informal}

\epspic{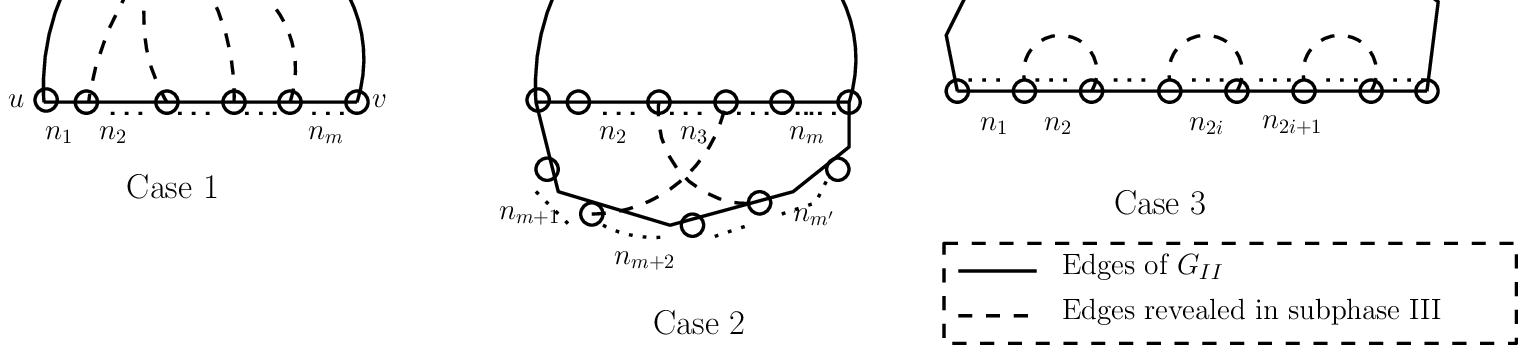}{}{Possible scenarios
during sub-phase III.}{fig:examples}

First we should note that the situation here is quite different from the
randomized case. In the randomized case the upper bound on the number of
faults is {\em not influenced} by the new edges revealed in sub-phase III.
In contrast, in  the deterministic case, the added edges can 
increase the number of faults. 
 For example, in case~1 in
Fig.~\ref{fig:examples}, at the end of sub-phase II we have a path $\la
u,\dots,v \ra$ in the vine decomposition $\Pi$ in $\GII$. Hence, the naive
lower bound for the number of faults in this vine is $g(\log \sum _i n_i)$,
whereas {\ddet} might have there almost $g(\sum _i \log n_i)$ faults, which
can be much higher. In this example, the solution is clear --- we should
construct a new vine decomposition that uses the new edges as part of the
backbone and has a value of $\Omega (\sum_i \log n_i)$. {\sl I.e.}, we
improve the lower bound on the number of faults of {\opt} to match the upper
bound.

Case~2 in Fig.~\ref{fig:examples}  is more complex. Here the construction of
a new vine decomposition is not obvious. The scenario addressed here
includes cases where new edges connect one $\GII$ path to another.
These new edges split the paths into sub-paths. If the lengths of the
resulting sub-paths were $(n_i)_i$, then one upper bound on the number of
faults for $\ddet$ during sub-phase III
would be $g \sum_i \log n_i$.
It is not obviously clear
that the adversary can actually force such a number of faults.
However, we will prove that in this scenario
it is possible to build a new vine decomposition with a value of $c \sum _i
\log n_i$ for some global constant $c>0$. Again, we have found matching
upper and lower bounds.

The situation becomes more complicated when the new edges do not cross path
boundaries, as in case~3 in Fig.~\ref{fig:examples}. In this case we can not
hope to construct a vine decomposition with value $\Omega(\sum _i \log
n_i)$, such a vine decomposition simply does not exist. Here we will have to
show that the upper bound on the number of faults for {\ddet} is indeed
$O(\log (\sum_i n_i) + \sum_i \log n_{2i})$, which is smaller than $\sum_i
\log n_i$.

The difference between cases 1,2 in Fig.~\ref{fig:examples} and case 3 is
that in cases 1 and 2 we used a simple upper bound on the number of faults
and could devise an appropriate vine decomposition for the lower bound on the
competitive ratio. In case 3 we need a more sophisticated upper bound as well
as  more involved construction of the vine decomposition for the lower bound.

\subsection{Preliminaries}
\seclab{preliminaries}

During sub-phase III, new pages might be requested (at most $g-1$ new pages).
As we can associate an amortized cost of $\Omega(1)$ to any offline algorithm
for every new page, we would like to ``ignore" them, but we need to consider
the connectivity relations they induce.

\begin{definition}
The \emph{simplification} of $\hG$ is a graph denoted by $G^s=(V,E)$, such
that $V$ is the set of vertices in $\GII$ and $E$ includes the edges of
$\GII$ \emph{and} edges $uv$ if there exists a path in $\GIII$ between $u,v
\in V$ such that all its internal vertices are \emph{not in $V$}, \ie, they
are new vertices requested during sub-phase III.
\end{definition}

It will be more convenient for us to work with $G^s$, as the set of vertices
in which we are interested (stale pages at the end of phase II) are already
in $G^s$, and $G^s$ has the same set of vertices as $\GII$, and just more
edges. However, in the conclusion  of the proof, we will have to reconsider
the fact that the actual graph, $\hG$, might have another $g-1$ vertices.

As mentioned in \secref{informal}, the vine decomposition of $\GII$ may not
give us a sufficiently high lower bound. In order to differentiate it from
the final vine-decomposition, we call it the {\em backbone bi-connected path
complex\/}  in $G^s$, or simply {\em the complex}.

Given a graph $G$ we denote its set of vertices by $V[G]$. For
$U\subseteq V[G]$ we denote the  sub-graph induced by $G$ on $U$
as $G|_U$. Given a simple path $p= \la v_1, \ldots, v_k \ra$ we define the
inner subpath $I(p)= \la v_2, v_3, \ldots, v_{k-1} \ra$.

\begin{definition}
\label{def:path} A {\em proper path} $p$ is a path in $G^s$ such that edges
with one endpoint in $V[I(p)]$ have their other endpoint in $V[p]$.
\end{definition}

Note that the new edges added to $\GII$ during the course of sub-phase III
decompose the paths of the complex into disjoint sub-paths. We view this
decomposition as an hierarchical process as follows:
\begin{enumerate}
\item
We ``add'' to the decomposition all the new edges that cross path
boundaries, which results in a decomposition to proper sub-paths.
\item
For every resulting sub-path we recursively construct a new decomposition.
\end{enumerate}
\noindent We now formally define the concepts \emph{decomposition} and
\emph{recursive decomposition}:

\begin{definition}
Given the complex $\C=(B,\Q)$, a {\em separating set} for $\C$ is a set $S$ of
vertices satisfying $S\subseteq \cup_{q\in \Q} V[q]$, and $\deg_{G^s}(v)\geq 3$ for
all $v\in S$.
\end{definition}

\begin{definition} \label{definition:decomposition-complex}
Given the complex $\C=(B,\Q)$ and a separating set $S$ for $\C$, we define the
{\em decomposition} $\D=(S, \PP)$ of $\C$ as follows:

Fix $q=\la v_1, v_2, \ldots, v_k \ra\in \Q$. Let $S\cap V[q] =
\{v_{i_1}, v_{i_2}, \ldots, v_{i_j}\}$, where $i_\ell >
i_{\ell-1}$, for $2\leq \ell \leq j$. Define the paths
\begin{eqnarray*} p_1 &=& \la v_1, v_2, \ldots, v_{i_1-1} \ra ,\\
    p_2 &=& \la v_{i_1+1}, v_{i_1+2}, \ldots, v_{i_2-1} \ra , \\
    &\vdots&\\
    p_j &=& \la v_{i_{j-1}+1}, v_{i_{j-1}+2}, \ldots, v_{i_j-1} \ra ,\\
    p_{j+1} &=& \la v_{i_j+1}, v_{i_j+2}, \ldots, v_k \ra .
  \end{eqnarray*}
Let $\PP(q)$ be the set of all the non-empty $p_\ell$ paths, $1\leq \ell \leq
j+1$. Let $\PP$ be the union of all $\PP(q)$, $q\in \Q$. $S$ is called
the separating set of $\D$, and it is denoted by $S[\D]$.

A {\em proper decomposition} $(S,\PP)$ is a decomposition in which all paths $p\in\PP$
are proper paths.
\end{definition}

\begin{definition} \label{def:proper-seperating}
Given a proper path $p= \la v_1,v_2 \ldots, v_k \ra$, a non-empty set
$S\subseteq V[p]$ is called a \emph{separating set} for $p$ if $S=
\{v_{i_1}, v_{i_2}, \ldots, v_{i_j}\}$, where $i_\ell>i_{\ell-1}$ for $2\leq
\ell\leq j$, satisfying:
\begin{enumerate}
\item For $1\leq \ell\leq j$:
\begin{enumerate}
\item  $\deg_{G^s}(v_{i_\ell})\geq 3$, or
\item $\deg_{G^s}(v_{i_\ell})=2$ and there is no edge between the sets $\{v_1, \ldots, v_{{i_\ell}-1}\}$ and
$\{v_{i_{\ell}+1}, \ldots, v_k\}$ in $G^s$. \end{enumerate}
\item For $1\leq \ell \leq j-1$, if $i_\ell<i_{\ell+1}-1$,
 then both $\deg_{G^s}(v_{i_\ell})\geq 3$ and $\deg_{G^s}(v_{i_{\ell+1}})\geq 3$.
\item If $i_1>1$ then $\deg_{G^s}(v_{i_1}) \geq 3$,
and if $i_j<k$ then $\deg_{G^s}(v_j)\geq 3$.
\end{enumerate}
\end{definition}

\begin{definition}\label{definition:decomposition-path}
Given a proper path $p=\la v_1,v_2 \ldots, v_k \ra$, and a separating set
for $p$, $S= \{v_{i_1}, v_{i_2}, \ldots, v_{i_j}\}$ ($i_1<\ldots< i_j$), the
\emph{decomposition} $\D=(S, \PP)$ of $p$ is defined as
follows:
Let $p_1, \ldots, p_{j+1}$ denote the paths
 \begin{eqnarray*} p_1 &=& \la v_1, v_2, \ldots, v_{i_1-1} \ra,\\
    p_2 &=& \la v_{i_1+1}, v_{i_1+2}, \ldots, v_{i_2-1} \ra , \\
    &\vdots&\\
    p_j &=& \la v_{i_{j-1}+1}, v_{i_{j-1}+2}, \ldots, v_{i_j-1} \ra ,\\
    p_{j+1} &=& \la v_{i_j+1}, v_{i_j+2}, \ldots, v_k \ra .
  \end{eqnarray*}
Let $\PP$ be the set of all the non-empty $p_\ell$ paths, $1\leq \ell \leq
j+1$. $S$ is also called the separating set of $\D$, and it is denoted by
$S[\D]$ or $S[p]$ if $\D$ is clear from the context.

A {\em proper decomposition} $(S,\PP)$ is a decomposition in which all paths $p\in\PP$
are proper paths.
\end{definition}

A key point in the above definitions is the allowance for vertices of degree
2 to be in the separating set of a decomposition of a proper path
(Def.~\ref{def:proper-seperating}), under certain restrictions. This solves
the problem imposed in the third example in Fig.~\ref{fig:examples}. The odd
subpaths there can be now part of the separating set, and not part of the
paths of the decomposition.

\begin{definition}
A \emph{recursive decomposition} $\D^R(q)$ of a proper path $q$ is a proper
decomposition $\D=(S, \PP )$ of  $q$, along with recursive decompositions
$\D^R(p)$, for each $p\in \PP$. We define the {\em value} of $\D^R(q)$
recursively as
\[ \tval(\D^R(q)) \defd \log |q|+\sum _{p \in \PP}
\tval(\D^R(p)).
\]
\end{definition}

\begin{definition}
A \emph{recursive decomposition $\D^R(\C)$ of the complex $\C=(B,\Q)$}  is a proper
decomposition $\D=(S, \PP )$ of $\C=(B,\Q)$ along with recursive
decompositions $\D^R(p)$, for each proper path $p \in \PP$. The {\em value}
of $\D^R(\C)$ is defined as \[ \tval(\D^R(\C))
\defd \sum_{p\in \PP} \tval(\D^R(p)). \]
\end{definition}

We use the shorthand $\D^R$ when the complex $\C$ or the proper path $q$ is
implicitly understood.

\begin{definition}
We define $\PP[\D^R]$ to be the set of all sub-paths in  the recursive
decomposition $\D^R$,
including sub-paths defined recursively. We also define \( \PP _g[\D^R] \defd
\{p \in \PP [\D^R] :\ |p| \geq g\} \), and
\[
\tval _g(\D^R)\defd \sum _{p \in \PP_g[\D^R]} \left(\log|p| -\log g \right ).
\]
\end{definition}

\begin{proposition}
\label{claim:v values} Let $\D^R=\D^R(\C)$ be a recursive decomposition of the
complex $\C=(B,\Q)$, and let $(S_0,\PP_0)$ be the top level proper
decomposition of $\C$ in $\D^R$. Then the following hold:
\begin{enumerate}
  \item \label{it:1} $\forall p\in\PP[\D^R]$,  $S_0 \cap S[p]=\emptyset$.
  \item $\forall p,q\in \PP[\D^R]$, if $p\neq q$ then $S[p] \cap S[q]= \emptyset$.
  \item $S_0 \cup\bigcup_{p\in \PP[\D^R]} S[p]=\bigcup_{q\in
  \Q}V[q]$.
  \item $\tval(\D^R)= \sum_{p \in \PP[\D^R]} \log |p|$.
  \item $\tval _1(\D^R) = \tval (\D^R)$.
  \item $|\PP(\D^R)|\leq \ell+t$, where $\ell=|\{v \in \bigcup_{q\in
  \Q}V[q]: \deg_{G^s}(v) \geq 3\}|$, and $t=|\Q|$.
\end{enumerate}
\end{proposition}
\begin{proof}
Items 1--5 follow immediately from the definitions. To prove 6 we first argue
that for any proper path $p$,  $|\PP(\D^R(p))|\leq \ell+1$, where $\ell$ is
the number of vertices of degree 3 or more in $p$. Next, we sum up over all
$p\in \PP_0$, getting an extra $t$.
\end{proof}

\begin{example}
\label{examp:simple decomp} Consider the following (maybe the simplest)
recursive decomposition of the complex $\C=(B,\Q)$: The first level consists
of a separating set that includes \emph{all} the vertices on the paths in
$\Q$ with degree at least 3. The rest of the vertices have degree 2 and are
grouped into sub-paths (which are proper). In the second level of the
recursive decomposition, each such sub-path is decomposed so that all its
vertices are in the separating set. As we shall prove in
Lemma~\ref{lemma:upper-bound}, any recursive decomposition of the complex,
implies an upper bound on $f_3$. However, the upper bound implied by this
recursive decomposition is not tight (see the discussion in
\secref{informal} about the third case in Fig.~\ref{fig:examples}), and we
need the full generality of the definition of recursive decomposition in
order to create a recursive decomposition that implies a tight upper bound
on $f_3$.
\end{example}

\subsection{The Proof}
\seclab{proof f_3}

\subsubsection*{The Upper Bound}

\begin{lemma}
\label{lemma:upper-bound} Let $\C=(B,\Q)$, $|\Q|=t$, be the complex in $G^s$
induced by {\ddet} at the end of sub-phase II. Let $\D^R$ be a recursive
decomposition of  the complex $\C$.
Let $g$ be the number of new pages in the phase, and $\ell$ be the number
of vertices in $\bigcup_{q\in \Q} q$ with degree at least 3 in $G^s$. Then,
\[ \hf= O \left (g(\ell+t+\tval _g(\D^R)) \right ). \]
\end{lemma}
\begin{proof}
There are at most $g$ faults not on the paths in $\Q$ ({\ie}, faults on new
vertices requested for the first time in the phase during sub-phase III).

We count the number of faults on the paths by charging faults to vertices of
degree $\geq 3$ in  $G^s$ or to paths in the recursive decomposition of the
complex.

If the fault is on a vertex of degree $\geq 3$ in $G^s$, then we charge it 
to the vertex. There are at most $\ell$ such faults.

Otherwise, the fault is on a vertex $v$ of degree 2 in $G^s$. From
Proposition~\ref{claim:v values}, $v$ must be in some $S[p]$ for some path
$p\in\PP(\D^R)$. It cannot be in the separating set of the complex itself
because all vertices in the separating set of the complex are of degree
$\geq 3$ in $G^s$.

We charge the path $p$ for the fault on $v$. We want to show that
there are at most $O(g \cdot \max\{\log |p| - \log g,1\})$ faults
associated with $p$. If this is true then
\begin{multline*}
f_3 \leq g+ \ell + \sum_{p\in \PP(\D^R), |p|<2g} (|p|-1)+ \sum _{p \in
\PP_{2g}(\D^R)}O(g \cdot (\log |p| - \log g)) \\ \leq g+\ell +( \ell+t) 2g +
O(g\tval _g(\D^R)).
\end{multline*}
and the proof of the lemma would be completed.

Let $S^2(p)$ denote the set of all vertices of degree $2$ in $S[p]$. Let
$U(p)\subseteq S^2(p)$ denote the set of unmarked vertices in $S^2(p)$. Over
time, when unmarked vertices are requested, they are removed from $U(p)$.

For $X\subseteq p$, let $C(X)$ denotes the minimal sized subpath of $p$ that
contains all vertices in $X$. We use the notation $C(U(p))$ to denote a set
whose size may decrease over time (as $U(p)$ itself is a set whose size may
decrease over time).

\begin{proposition} All vertices in $C(U(p))$ are unmarked.
\end{proposition}
\begin{proof}
The proof follows from the fact that $p$ is a proper path. Let $u,v\in
S^2(p)$, $u\neq v$, be  arbitrary distinct vertices  in $S^2(p)$. Assume
that $w\in C(\{u,v\})$ and $x\notin C(\{u,v\})$. Then any path from $x$ to
$w$ must pass through either $u$ or $v$. Thus, if any vertex in $C(\{u,v\})$
was requested this implies that either $u$ or $v$ was requested, {\sl i.e.},
either $u$ or $v$ is marked.

Taking $u$ and $v$ to be the extreme points of $C(U(p))$ (which must also be
in $S^2(p)$) concludes the proof of the claim.
\end{proof}

Let $q\in \Q$ be a path in $\C$ that contains $p$ as a sub-path. Let
$M(U(p))$ denote the longest unmarked subpath of $q\in \mathcal{Q}$
containing $C(U(p))$, so $M(U(p))$ also varies over time. When {\ddet} evicts
a vertex from $M(U(p))$, it is the closest stale page to the midpoint of
$M(U(p))$. As there are at most $g$ non-stale pages in $M(U(p))$, the evicted
page is at distance at most $g$ from the midpoint of $M(U(p))$.

\begin{proposition}
After $g+1$ evictions from $S^2(p)$, $|M(U(p))|\leq 2(|p|+g)$
\end{proposition}
\begin{proof}
We first claim that after any fault in $S^2(p)$, $p$ includes at least one
of the extreme points of $M(U(p))$. This follows because $C(U(p))\subseteq
M(U(p)) \subseteq q$, $C(U(p))\subseteq p \subseteq q$, so $p \cap M(U(p))
\neq \emptyset$. As $p$ is a subpath of $q$ and $M(U(p))$ is a subpath of
$q$, either $p$ is a subpath of $M(U(p))$ or an extreme point of $M(U(p))$
is in $p$. $p$ cannot be a subpath of $M(U(p))$ because $p$ contains a
marked vertex whereas $M(U(p))$ consists only of unmarked vertices. Thus, it
must be that an extreme point of $M(U(p))$ is in $p$.

There must be at least one fault in $S^2(p)$ prior to the ($g+1$) eviction
from $S^2(p)$. On the next eviction from $S^2(p)$ following the first fault
from $S^2(p)$, we know a vertex of distance at most $g$ from the midpoint of
$M(U(p))$ is in $S^2(p)\subseteq p$, whereas one endpoint of $M(U(p))$ is in
$p$, thus $|M(U(p))| \leq 2(|p|+g)$.
\end{proof}

The $g+1$ evictions from $S^2(p)$ described in the lemma above can cause at
most $g+1$ faults in $S^2(p)$. This means that we have associated at most
$g+1$ faults with $p$ prior to the configuration where $|M(U(p))| \leq
2(|p|+g)$.

We now count the number of evictions from this point onwards, this is  a
bound on the number of faults in $S^2(p)$.

After every $g$ evictions, the size of $M(U(p))$ decreases by a factor of
roughly 1/2. So long as $|M(U(p))|\geq 10 g$, this factor is at most $6/10$.
Thus, after $O(g (\log |p| - \log (10 g)))$ evictions we have $|M(U(p))|\leq
10 g$. On the remaining vertices we can fault at most $10 g$ times, giving
us a total number of faults on this stage of $O(g \cdot \max\{\log |p| -
\log g,1\})$.
\end{proof}

\subsubsection*{The Lower Bound}

The idea is to construct a vine decomposition $\V=(B,\Q)$ on $k+1$ vertices
(see \defref{vine-decomp}) whose value matches the value of \emph{some}
recursive decomposition $\D^R$, up to a constant factor:
 \begin{equation} \label{eq:lb:goal}
 \vval(\V)\geq \Omega(1)\cdot \tval(\D^R).
 \end{equation}

If this is true, then from Lemma~\ref{lemma:upper-bound}, \lemref{tree-lb},
\propref{leaves}, and \lemref{vinedecomp-lb} it follows that the competitive
ratio that the adversary can force upon any online algorithm is no worse
than the competitive ratio of $\ddet$.

Consider a decomposition $\D=(S,\PP)$ (this decomposition is either a
decomposition of a proper path or of a complex). As a first step towards
obtaining our goal of the previous paragraph, we seek a vine-decomposition
$\V=(B,\PP')$ such that
\begin{enumerate}
\item The set of paths in $\PP'$ is a subset of the set of paths in $\PP$.
\item The backbone of the vine decomposition $\V$ includes the
separating set $S$ of $\D$ and includes the paths of $\PP \setminus \PP'$.
\item The value of $\V$ is  no less than a constant fraction of the value of $\D$.
\end{enumerate}

\begin{definition} \label{def:proper-vd}
\label{def:vine proper} A {\em proper vine decomposition} $\V$ of a {\em
proper path} $p$
is a vine decomposition
of the graph induced by $G^s$ on the vertices of $p$,
such that the paths of $\V$ are sub-paths of $p$,
and the endpoints of $p$ are in the backbone of $\V$.
\end{definition}

\begin{definition}
\label{def:gvs} A {\em \gvs} of a decomposition $\D=(S, \PP )$ of a proper
path $q$ is a set $A(\D)=\PP' \subseteq \PP $ (called vines) such that
\begin{itemize}
\item
The induced graph on $S \cup \bigcup _{p \in \PP \setminus A(\D)} V[p]$ is
connected.
\item
  \(\sum _{p\in A(\D))} \log |p| \geq c \sum_{p \in \PP} \log |p| \),
       where $c=1/32$.
\item The endpoints of $q$ are
      in $S\cup \bigcup _{p \in \PP \setminus A(\D)} V[p]$.
\end{itemize}

It follows that $(S \cup\bigcup_{p \in (\PP \setminus A(\D))} V[p] ,\,
A(\D))$ is a proper vine decomposition of $q$.
\end{definition}

\begin{definition}
A {\em \gvs} of a decomposition $\D=(S, \PP )$ of a complex $\C=(B,\Q=\{q_1,
\ldots, q_t\})$ is a set $A(\D)=\PP'\subseteq \PP$ such that
\begin{itemize}
\item
The induced graph on $\bigcup _{p \in \PP \setminus A(\D)} V[p]\bigcup B
\bigcup S$ is connected.
\item
  \(\sum _{p\in A(\D)} \log |p| \geq c \sum_{p \in \PP} \log |p| \),
       where $c=1/32$.
\end{itemize}

It follows that $(B\cup S\cup \bigcup_{p \in (\PP \setminus A(\D))} V[p] ,\,
A(\D))$ is a vine decomposition of $G^s$.
\end{definition}

To construct a vine decomposition as required in \eqref{eq:lb:goal},
we make use of a special type of decomposition, called an \emph{irreducible
decomposition}.

\begin{definition}
Given a path $p=\la v_1,\ldots,v_m \ra$, denote the {\em span} of an edge
$e=v_i v_j$ between two vertices in $p$ by $|e|_p\defd |j-i|$. An {\em
irreducible path} $p$ is a  proper path in which for every edge $e$ whose
endpoints are in $p$, $|e|_p< |p|^{1/4}$. An {\em irreducible decomposition}
is a proper decomposition $\D=(S,\PP)$ such that all paths $p\in \PP$ are
irreducible.
\end{definition}

The following Lemma shows that it is possible to construct an irreducible
decomposition along with a corresponding {\gvs}.

Given a simple path $p=\la v_1,v_2, \ldots, v_k \ra$, a maximal subpath of
degree-2 vertices in $p$ is a subpath $q=\la v_i, v_{i+1}, \ldots, v_j \ra$,
$1\leq i \leq j \leq k$, such that the $\deg_{G^s}(v_\ell) = 2$ for all
$i\leq \ell \leq j$ while if $i>1$ then $\deg_{G^s}(v_{i-1})>2$, and if
$j<k$ then $\deg_{G^s}(v_{j+1})>2$.

\begin{lemma}
\label{lemma:full vine selection}\
\begin{enumerate}
\item
Given a proper path $q$ and assuming that every maximal subpath of degree-2
vertices has at least $15$ vertices, then $q$ has an irreducible
decomposition $\D=(S,\PP )$ and a corresponding {\gvs} $A(\D)$.
\item
Given a complex $\C=(B,\Q)$, and assuming that for all  $q\in \Q$ every
maximal subpath of degree-2 vertices of $q$ has at least $15$ vertices, then
$\C$ has an irreducible decomposition $\D=(S,\PP )$ and a corresponding
{\gvs} $A(\D)$.
\end{enumerate}
\end{lemma}
The proof of the lemma appears in \secref{combinatorial}.

We are now ready to construct the required recursive decomposition and vine
decomposition, as required in Equation \eqref{eq:lb:goal}. We give a
constructive algorithm that builds both simultaneously, the algorithm makes
use of recursive decompositions.

\begin{enumerate}
\item Use Lemma~\ref{lemma:full vine selection} to obtain an
irreducible decomposition $\D=(S, \PP )$ and related vine selection $A(\D)$;
\item Recursively find for every sub-path $p\in \PP$, a vine decomposition
      and recursive decomposition;
\item  Using the resulting recursively obtained vine decompositions
and the vine selection $A(\D)$, we construct the required vine decomposition
(Lemma~\ref{lemma:inductive gvd} in \secref{combinatorial}).
\end{enumerate}

Lemma~\ref{lemma:gvd} summarizes the construction for proper paths, the
construction for a complex is handled in Lemma~\ref{coroll:matching}.

\begin{lemma}
\label{lemma:gvd} $\exists c'$, $c\geq c'>0$, such that any proper path $q$,
with each maximal subpath of degree-2 vertices having at least $15$ vertices,
has a recursive decomposition $\D^R$ along with a proper vine decomposition
$\V$ of $q$ such that \( \pval(\V) \geq c' \cdot [ \tval(\D^R) - \log |q|
]\).
\end{lemma}

\noindent The proof appears in \secref{combinatorial}. Lemma~\ref{lemma:gvd}
is used in the following lemma to construct the vine-decomposition for the complex.

\begin{lemma}
\label{coroll:matching} Given the complex $\C=(B,\Q=\{q_1, \ldots, q_t\})$ in
$G^s$, let $\ell$ be the number of vertices  $x\in\bigcup_{q \in \Q}
V[q]$ such that $\deg_{G^s}(x)\geq 3$. There exists a recursive decomposition
$\D^R=\D^R(C)$ and a vine-decomposition $\V$ of $\C$ such that \( \max \{ \ell+t,
\vval(\V) \} = \Omega(\ell+t+ \tval(\D^R)) .\)
\end{lemma}
\begin{proof}
First we change $G^s$ by adding new degree-2 vertices in such a way that
every maximal subpath of degree-2 vertices is of length at least 15 (we
need at most $14(l+t)$ new vertices). Denote the resulting graph $G'$. The complex
$\C$ in $G^s$ naturally induces a complex $\C'=(B,\Q'=\{q'_1, \ldots,
q'_t\})$ in $G'$.
 From Lemma~\ref{lemma:full vine selection} we have an  irreducible
decomposition $\D'=(S,\PP)$ of the complex, along with a {\gvs} $A(\D')$. The
vine selection $A(\D')$ induces a vine decomposition, $\V'_1$,  on $G'$ such
that $\pval(\V'_1) \geq c \sum_{p\in \PP} \log |p|$.

 From Lemma~\ref{lemma:gvd}, we have for each path $p \in \PP$, a recursive
decomposition $\D^R(p)$ and a proper vine decomposition $\V'_2(p)$ such that
$\pval(\V'_2(p)) \geq c' [ \tval(\D^R(p)) - \log |p|]$.

We construct a new vine decomposition $\V'_2$ for $G'$. The set of paths in
$\V'_2$ is the union of the paths in all the $\bigcup_{p\in \PP}\V'_2(p)$.
The backbone of $\V'_2$ is the union of the sets $B$, $\{B'_p\}_{p\in \PP}$, and
$S$, where $B$ is the backbone of $\C'$, $B'_p$ is the backbone of the
proper vine decomposition of $\V'_2(p)$, $p \in \PP$, and $S$ is the
separating set of $\D'$.

We show that $\V'_2$ is indeed a vine decomposition by showing  that the
backbone is connected and that all the paths of $\V'_2$ are adjacent to the
backbone at their endpoints.

Consider a path $q=\la v_1,\ldots,v_r \ra \in \Q$. We shall see that if 
$q\cap
S$ is connected to $B$ in the backbone of $\V'_2$, whenever $q\cap S$ is non-empty. 
Let $q \cap
S=\{v_{i_1},\ldots, v_{i_s}\}$, ordered in their order on $q$. Let $i_0=0$,
and $v_{i_0} \in B$ a vertex in $B$ adjacent to $v_1$. We prove that for all
$j\in\{0,\ldots,s-1\}$, $v_{i_j}$ and $v_{i_{j+1}}$ are connected in the
backbone of $\V'_2$. Note that either $i_j+1=i_{j+1}$, and in this case
those two vertices are adjacent in $q$, or otherwise there is $p\in\PP$ a
path of $\D'$ between them. As $\V'_2(p)$ is a proper vine decomposition of
$p$ (see Def.~\ref{def:proper-vd}), the endpoints of $p$ are in $B'_p$ and
thus connected in $B'_p$. So $v_{i_j}$ and $v_{i_{j+1}}$ are also connected
via $B'_p$.

Within every path $p\in \PP$ we have a valid vine decomposition, whose
backbone is connected to $B$ via vertices in $S$, thus $\V'_2$ is a legal
vine decomposition. It also follows that $\pval(\V'_2) = \sum_{p\in \PP}
\pval(\V'_2(p))$.

Let $\V' = \V'_1$ if $\pval(\V'_1)\geq \pval(\V'_2)$, and let $\V'=\V'_2$
otherwise. Let $\D'{}^R$ be a recursive decomposition of
$\C'$ defined in a natural way as $\D'{}^R=(S,\{ \D^R(p):\ p\in \PP \})$.

\begin{align*}
\vval(\V')  &  =
  \max\{ \pval(\V'_1), \pval (\V'_2)\}
  \geq \tfrac{1}{2} \pval(\V'_1) +\tfrac{1}{2} \pval(\V'_2) \\
& \geq  \tfrac{1}{2} c \sum_{p \in \PP } \log |p|  +
    \tfrac{1}{2} c'  \sum_{p \in \PP } \bigl [ \tval(\D^R(p)) -
    \log |p|  \bigr ] \\
& = \tfrac{1}{2} c' \sum_{p \in \PP } \tval(\D^R(p))  = \tfrac{c'}{2}
\cdot \tval(\D'{}^R)
\end{align*}

Define $\D^R$ and $\V$ on $G^s$, by removing from $\D'{}^R$ and $\V'$ the
artificial vertices. $\D^R$ remains a recursive decomposition of $\C$, $\V$
remains a vine decomposition of $G^s$, and $\tval(\D'{}^R) \geq \tval(\D^R)$.
There are at most $\ell+t$ vines in $\V'$ and the value of every each is
reduced by at most $\log 16=4$, so $\vval(\V) \geq \vval(\V') - 4(\ell+t) $.
Therefore
\[
5(\ell+t) + \vval(\V)  \geq (\ell+t) +\vval(\V') \geq \ell+t
+\tfrac{c'}{2}\tval(\D'{}^R)
         = \Omega(\ell+t+ \tval(\D^R)).
 \quad \quad \qed \]
\renewcommand{\qed}{} \end{proof}

The vine decomposition $\V$ in Lemma~\ref{coroll:matching} is of $G^s$,
instead of $\GIII$, and furthermore, might have more than $k+1$ vertices.
Thus we are not quite done yet. The following lemma allows us to reduce the
number of vertices in $\V$ to $k+1$, or to find a ``big cycle" in it. The
proof appears in \secref{combinatorial}.

\begin{lemma} \label{lem:throw-h}
Given a complex $\C=(B_1,\Q)$ on the vertex set $V$ and a vine decomposition
$\V=(B_2,\PP)$ of $\C$ and an
integer $h>0$, one of the following holds:
\begin{enumerate}
\item $\max_{q\in \Q} |q| \leq 8(h+1)$.
\item $\exists p \in \PP$, such that $\log |p| \geq \vval(\V)/24$.
\item There exists a set of vertices $T\subseteq \cup_{q\in \Q} q$
and vine decomposition $\V'$ on the vertex set $V\setminus T$, such that
$|T|\geq h$ and $\vval(\V') \geq \vval(\V)/24$. Furthermore, for each $q\in
\Q$, $T\cap q$ is a subpath of $q$, and at least one of its endpoints is
adjacent to the backbone of $\V'$.
\end{enumerate}
\end{lemma}

We are ready to conclude \lemref{bounded-f3}.

\begin{proof}[Proof of \lemref{bounded-f3}]
At the end of sub-phase II, $(B,\overline{\Pi})$ is a vine
decomposition. $(B,\overline{\Pi})$ induces a complex $(B,
\{q_1,\dots,q_{t}\})$ on the simplification graph $G^s$.

Let $\D^R=\D^R(\C)$  be the recursive decomposition $\C=(B,Q)$ and $\V$ the
vine decomposition of $\C$ obtained by Lemma~\ref{coroll:matching}. From
Lemma~\ref{lemma:upper-bound} we know that $\hf = O(g (\ell+t+
\tval_g(\D^R)))$. Hence, we are left to prove that $\ell+t+ \tval_g(\D^R)=
O(\compr^\infty(\hG,k))$.

First we observe that $\ell=O(\compr^\infty(\hG,k))$. This is true since by
\propref{leaves}, there exists a tree on $k+1$ vertices with $\Omega(\ell)$
leaves, so by \lemref{tree-lb}, $\ell=O(r^\infty(\hG,k))$.

Next, we observe that $t=O(\compr^\infty(\hG,k))$. Indeed, when
disconnecting the paths $q_1,\dots,q_t$ at their midpoints, and by removing
$g-1$ vertices from these paths, we get a subgraph on $k+1$ vertices with at
least $t/2$ leaves, so $t=O(r^\infty(\hG,k))$. 
(Note that the case $\sum_i q_i <2g$ is easy).

We are left to prove that $\tval_g(\D^R)=O(r^\infty(\hG,k))$. If
$\vval(\V)\leq \ell+t$, then by Lemma~\ref{coroll:matching}
$\tval_g(\D^R)\leq \tval(\D^R)=O(\ell+t)=O(r^\infty(\hG,k))$, and we are
done. Otherwise,
 we apply Lemma~\ref{lem:throw-h} with $h=g-1$ on the
vine decomposition $\V$ in the complex $\C'$. One of the following must
happen.

\begin{enumerate}
\item If $\max_{q\in\Q}|q| \leq 8h$, then $\tval_g(\D^R)\leq |\PP_g[\D^R]|
(\log 8g -\log g)=O(\ell+t)= O(r^\infty(\hG,k))$.

\item If $\exists p \in \PP$ such that $\log |p| \geq \vval(\V)/24$, then
let $q_1\in \Q$ be the longest path in $\C$, so $|q_1|\geq |p|$, and
  \( \log |q_1| \geq \log |p| \geq \vval(\V)/24 \geq d \tval(\D^R), \)
for some global constant $d>0$.

If $|\PP_g(\D^R)| \geq \frac{1}{d}$ then
  \[
   \tval _g(\D^R) \leq \tval(\D^R)- \bigl |\PP _g(\D^R) \bigr | \cdot \log g
   \leq \tval(\D^R)-  \tfrac{1}{d} \cdot \log g \leq \tfrac{1}{d}(\log
   |q_1| - \log g).
\]
If, on the other hand, $|\PP _g(\D^R)| < \frac{1}{d}$, then
\begin{align*}
 \tval_g(\D^R) 
 = \!\! \sum_{p \in {\PP}_g(\!\D^R)} \!\!
   (\log |p| - \log g)  \leq \tfrac{1}{d} (\log |q_1| - \log g).
\end{align*}
In either case, $\tval_g(\D^R) =O( \log|q_1| - \log g )$. From
\lemref{bigcycle-lb}, $\log |q_1| - \log g = O(\compr^\infty(\hG,k))$, thus
$\tval_g(\D^R)= O(\compr^\infty(\hG,k)))$.

\item There exists $T$,  a set of vertices in $G^s$, $|T|\geq g-1$,
and a vine decomposition $\V'$ on the set of vertices of $G^s\setminus T$ such
that $\vval(\V') \geq \vval(\V)/24$. $\V'$ is a vine decomposition in $G^s$.
We transform it to a vine decomposition in $\GIII$ as follows. First, each
edge of the backbone of $\V'$ that does not appear in $\GIII$ is replaced with
a path of vertices in $V[\GIII]\setminus V[\GII]$ that realizes this edge.
The resulting vine decomposition still has at most $k+1$ vertices. Next we
augment this vine decomposition to have exactly $k+1$ vertices, by adding
vertices removed from $\V'$, back to the backbone. This is done by adding
vertices from $T$ to the backbone in the exact amount to reach $k+1$ in the
vine-decomposition. Since the vertices of $T$ form subpaths in the paths of
the complex $\C$ that at least one of their endpoints adjacent to the
backbone of $\V'$, we can add them in such away that the augmented backbone
remains connected.

The resulting vine decomposition $\V''$ has the same value as $\V'$. From
\lemref{vinedecomp-lb}, $\vval(\V'')=O(\compr^\infty(\hG,k))$. Thus,
$\tval_g(\D^R)\leq \tval(\D^R)= O(\vval(\V))=O(\vval(\V''))
=O(\compr^\infty(\hG,k))$. \qed
\end{enumerate}
\renewcommand{\qed}{}
\end{proof}

\subsection{Proofs of the Combinatorial Lemmas} \seclab{combinatorial}

In this section we supply the proofs omitted from the previous section.

\subsubsection*{Proof of Lemma~\ref{lemma:full vine selection}}

Constructing the irreducible decomposition is done using {\em refinements}.
\begin{definition}
A {\em refinement} of a decomposition $\D=(S,\PP)$ is a decomposition
$\D'=(S', \PP')$ of the same object such that $S \subsetneq S'$, and the
paths in $\PP'$ are sub-paths of the paths in $\PP$.
\end{definition}

The next two lemmas show how to construct an irreducible decomposition along
with a corresponding {\gvs}.
\begin{lemma}\label{lemma:decompose-proper}
For every proper path $p$ there exists a
decomposition $\D=(S, \PP)$, and a vine selection $A(\D)$, such that $S$
includes the endpoints of $p$.\footnote{$\D$ is not necessarily a proper
decomposition. Properness will be dealt in Lemma~\ref{lemma:vine-selection}.}
\end{lemma}
\begin{proof}
Let $p=\la v_1,\ldots,v_{|p|-1} \ra$. $S$ is constructed in two stages. A
path-edge $v_i v_{i+1}$ is called ``covered" if there exists an edge $e=v_j
v_l$ such that $j\leq i<i+1\leq l$ and $\{j,l\}\neq \{i,i+1\}$. Remove the
uncovered edges in $p$, and consider the resulting connected components of
the graph induced on $p$. Note that these connected components are sub-paths
of $p$. Singletons are put in $S$. We consider each non-singleton sub-path
individually, finding a decomposition and a vine-selection for each.
Combining the decompositions and vine selections for the sub paths gives the
required result. Note that each such  subpath is a proper subpath and any of
its  path-edges is ``covered" by an edge whose endpoints are in that
sub-path.

 From now on, assume $p'=\la v_1,\dots,v_m \ra$ is a proper path, whose
path-edges are all covered by some edge whose endpoints are in $V[p']$.

We choose a set of edges $E'=\{e_1,\dots,e_{m'}\}$ by induction as follows:
let $l_1=1$, $e_1=v_{l_1} v_{r_1}$ for the maximum available $r_1$. Assume
inductively that $e_1=v_{l_1} v_{r_1},\dots,e_{i-1}=v_{l_{i-1}}
v_{r_{i-1}}$, where $r_{i-1}<m$, were already chosen. Consider the edge
$e=v_l v_r$ such that $l\leq r_{i-1}$ and $r$ is maximized under this
condition. Note that $r>r_{i-1}$ since otherwise $v_{r_{i-1}} v_{r_{i-1}+1}$
is not covered. We consider two cases:
\begin{enumerate}
\item
If $r_{i-1} -l\leq r - r_{i-1}$, then we set $e_i=e$.
\item Otherwise, consider the edge $e'=w_{l'} w_{r'}$ such that $l' \leq
r$ and $r'$ is maximized under this condition. Note that $r'>r$. We set
$e_i=e'$.
\end{enumerate}

We continue until $v_{m}$ is reached, and let $r_{m'}=m$. It is easily
checked that
 $\forall i\ ( 1\leq i \leq m')$, $r_{i-1} < l_{i+1}$ (taking $r_0=1$,
 $l_{m'+1}=m$),
and that both $(l_i)_i$ and $(r_i)_i$ are strictly increasing sequences.

\epspic{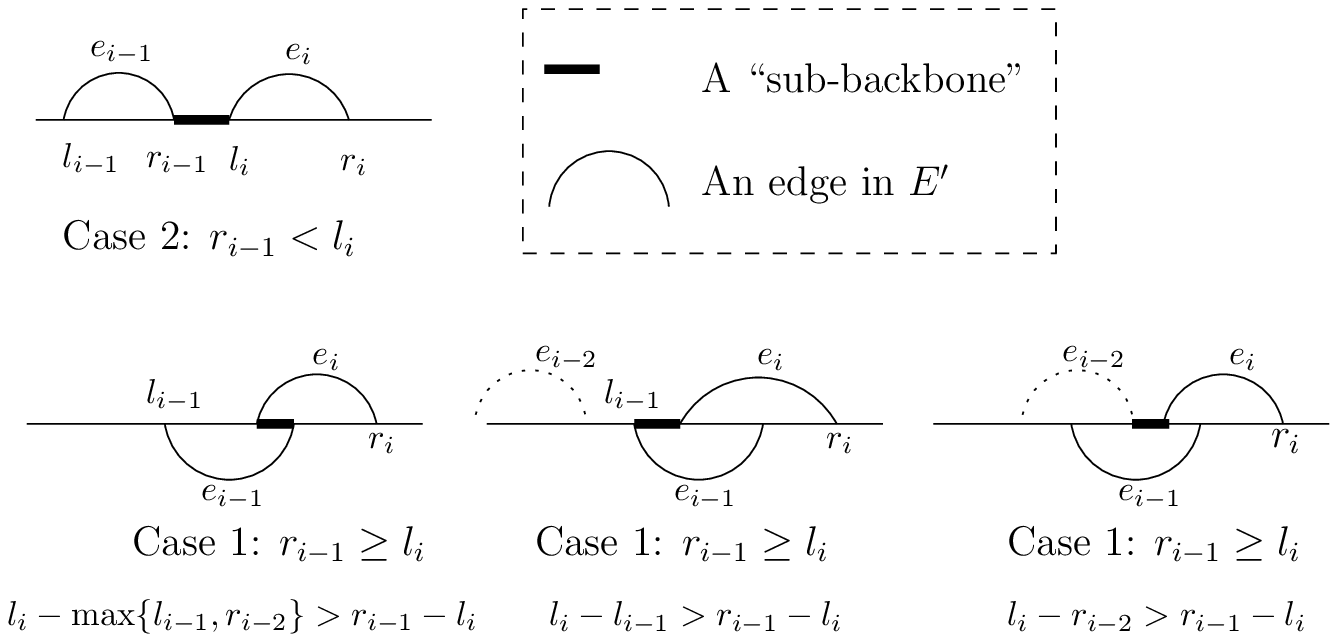}{} {The different cases in the proof of
Lemma~\ref{lemma:decompose-proper}. In case~1 the ``sub-backbone" path is
chosen to be smaller than an adjacent vine. In case~2 the
``sub-backbone" path is chosen to be $\la v_{r_{i-1}+1}, v_{l_i-1}\ra$, which
is smaller than $\la v_{\max\{r_{i-2},l_{i-1}\}+1},\ldots v_{r_{i-1}}\ra$.
}{fig:proper selection}

We construct $\D$ by adding to $S$ the endpoints of the edges in $E'$. Let
$p'_1,\dots,p'_{s'}$ denote the resulting decomposed sub-paths, in their
order in $p'$. The vine selection $A(\D)$ is constructed by taking roughly
half of the sub-paths as vines, as described below (see also
Fig.~\ref{fig:proper selection}). For $2 \leq i \leq m'$:

\begin{enumerate}
\item \label{en:2}
If $r_{i-1}\geq l_i$, let $h=1+ \max\{l_{i-1}, r_{i-2}\}$.
We declare the shorter of the two sub-paths $\la
v_{h},\dots, v_{l_i -1} \ra$ and
$\la v_{l_i +1},\dots, v_{r_{i-1} -1} \ra$ as a
``sub-backbone", {\ie}, it will not be part of $A(\D)$.
\item \label{en:1}
If $r_{i-1}<l_i$, we declare the sub-path $\la
v_{r_{i-1}+1},\dots,w_{l_i -1} \ra$ as ``a sub-backbone".
\end{enumerate}

$A(\D)$ consists of all the other sub-paths, not declared as
``sub-backbones". It is easily checked that the sub graph induced on $S \cup
\bigcup_{p'_i \not \in A(\D)} p'_i$ is  connected and adjacent to the
endpoints of $p'_i$, $p'_i\in A(\D)$.

In case~(\ref{en:2}) The sub-path not declared as sub-backbone is in $A(\D)$
and is bigger. In case~(\ref{en:1}), the sub-path $\la v_{h},\dots,
v_{r_{i-1}-1} \ra$ is in $A(\D)$, and is  bigger than the sub-backbone $\la
v_{r_{i-1}+1},\dots,v_{l_i -1}\ra $. Hence every sub-path not selected to
$A(\D)$, has an adjacent sub-path in $A(\D)$ which is bigger. This means
that we have constructed a mapping $f:\{ p'_1,p'_2,\ldots,p'_{s'} \}
\rightarrow A(\D)$, such that $|f^{-1}(\{p'_i\})| \leq 3$ for all $p'_i \in
A(\D)$, and $|p'_i| \leq |f(p'_i)|$ for all $i\in \{1,\ldots,s' \}$.
Therefore
\[ \sum_{p'_i \in A(\D)} \log (|p'_i|+1) \geq \frac{1}{3} \sum_{i=1}^{s'} \log
(|p'_i|+1).
\]
\end{proof}

\begin{lemma}
\label{lemma:vine-selection} Given an initial decomposition
$\D^I=(S^I,\PP^I)$, in which
every maximal path of degree-two vertices in $\PP^I$
contains at least 15 vertices, and a vine selection $A(\D^I)$.
Then there exists a refinement irreducible decomposition $\D=(S,\PP)$ and a
vine selection $A(\D)$.
\end{lemma}
\textbf{Note}: The lemma holds, both for a decomposition of the complex, and
for a decomposition of a proper path. We denote by $B$ the backbone of the
complex and we assume $B=\emptyset$ for a decomposition of a proper path.
\begin{proof}
The proof is by induction. Given a decomposition $\D=(S,\PP)$ and a vine
selection $A(\D)$ such that $\sum_{p\in A(\D)} \log |p| \geq c\sum_{p \in
\PP} \log |p|$, if $\D$ is an irreducible decomposition then we are done.
Otherwise, there exists an edge $e=u v$ that contradicts $\D$ being irreducible. 
We call $e$  a violating edge.

In this case we build a refinement $\D'=(S',\PP')$ from $\D$ as follows.
\begin{itemize}
 \item $S'=S \cup (\{u, v \} \setminus B)$.
 \item If $u, v\in p$, for some $p\in \PP$, we decompose $p$
to three sub-paths: $p_1$, $p_2$ and $p_3$. Then define $\PP' \defd \PP \cup
\{p_1, p_2, p_3 \} \setminus \{p\}$.
 \item If $u\in p,\ v\in q$, such that $p, q \in \PP$ and $p \neq q$,
      we decompose $p$ to two subpaths $p_1$ and $p_2$
       and $q$ to two sub-paths $q_1$, and $q_2$.
      Then define
$\PP' \defd \PP  \cup \{p_1, p_2, q_1, q_2 \}\setminus \{p, q\}$.

\end{itemize}
$\D'$ is obviously a legal decomposition. Moreover, the set of violating
edges decreased (with $e$ becoming a non-violating edge).
Therefore, this process is finite, and at the end we get an irreducible
decomposition.

We are left to find a vine selection $A(\D')$ for $\D'$ such that
\begin{equation}
\label{equ:equ1} \sum_{p'\in A(\D')} \log |p'| \geq c\sum_{p'\in \PP'} \log
|p'|.
\end{equation}
Denote
\begin{align*}
\tilde{\Delta} & \defd \quad \sum_{p'\in \PP'} \log |p'| -\sum_{p \in \PP}
\log |p|,\\ \Delta  & \defd \sum_{p'\in A(\D')}\!\!\!\!\! \log |p'|  -
\!\!\!\! \sum_{p\in A(\D)} \!\!\!\!\! \log |p|.
\end{align*}
Note that $\tilde{\Delta} \geq 0$. We will see how to choose $A(\D')$ such
that $\Delta \geq c \tilde{\Delta}$ (proving equation~(\ref{equ:equ1})), and
the sub graph spanned by $B \cup S' \cup \bigcup _{p' \not \in A(\D')} p'$
remains connected.

We denote by $B(\D)$ the subgraph induced on
$B\cup S \cup \bigcup_{p \in \PP \setminus A(\D)} p$,
and $B(\D')$ the subgraph induced on
$B \cup S' \cup \bigcup _{p'  \in \PP' \setminus
A(\D')} p'$. \emph{I.e.}, $B(\D)$ is the backbone of the vine decomposition
in the old decomposition, and $B(\D')$ is the backbone of the vine
decomposition after the refinement.

We use the notation $n_i=|p'_i|$. From the assumptions of the lemma, $n_i
\geq 16$, and we make use of fact that for $ n \geq 16$,  $\log n -2 \geq
\tfrac{1}{2} \log n$.

We do a case analysis according to the places of $u$ and $v$. The different
cases are also illustrated in Fig.~\ref{fig:vines selection}.
\epspic{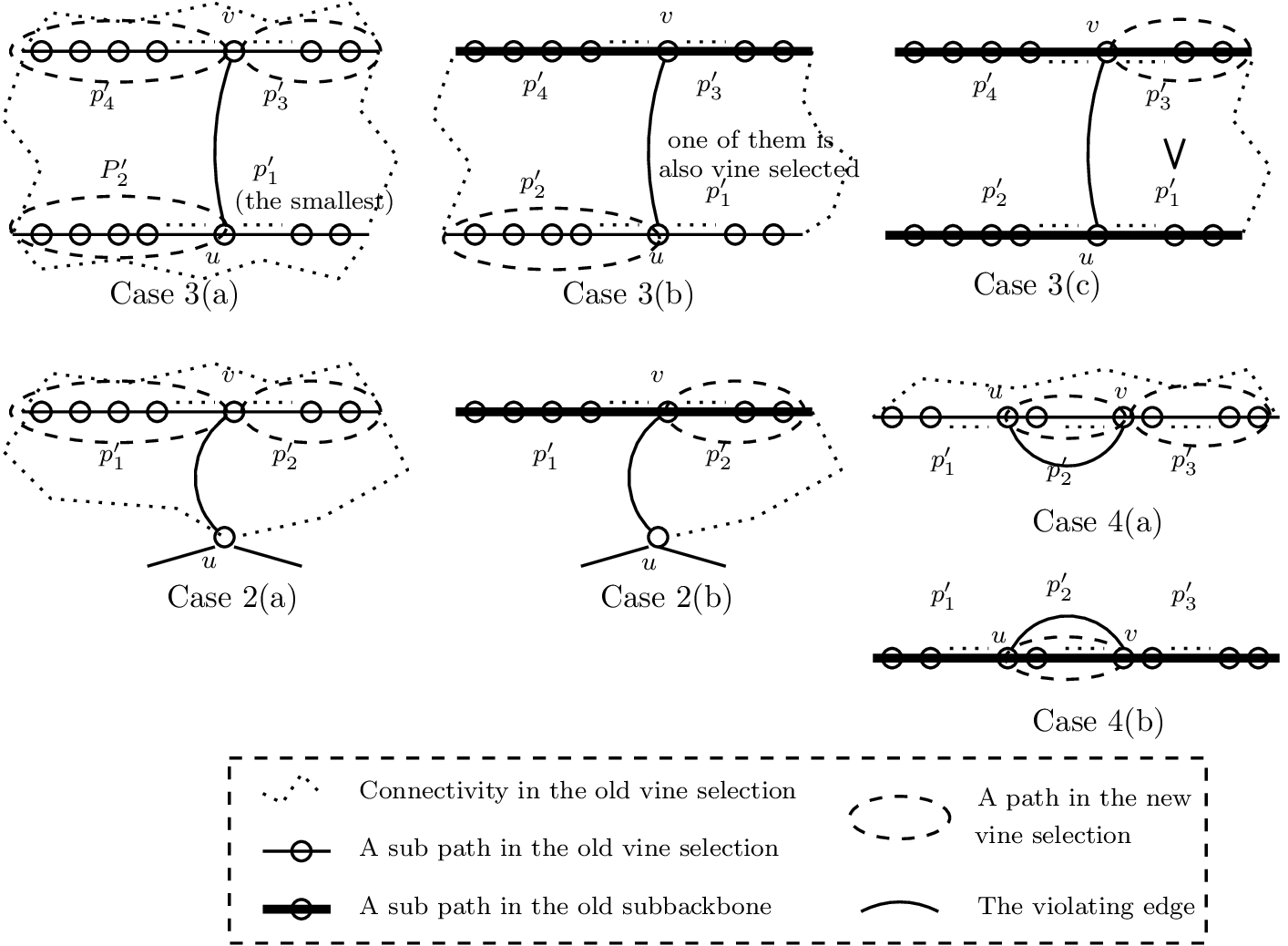}{}
 {The different cases in Lemma~\ref{lemma:vine-selection}.
  In case 2, the edge $e=u v$ is between a path and the backbone.
  In case 3, $e$ is between two different paths.
  In case 4, $e$ is inside the same path. The different sub-cases
  corresponds to the different possibilities of the paths to be either
  in the vine-selection, or as sub-backbones.
} {fig:vines selection}

\begin{enumerate}
\item Both $u$ and $v$ are not on paths in $\D$
(they are in $S \cup B$). In this case $e$ is not a violating edge, which is
impossible.

\item One of $u$ or $v$ is on a path in $\D$ and the other is not.
In this case $|\PP'|=|\PP|+1$. Assume $v\in p$, for some $p \in \PP$, and
$u\in S \cup B$, so $p=p'_1 \cup \{v \} \cup p'_2$, where $p'_1,p'_2 \in
\PP'$. As $|p|=n_1+n_2$ (since $| \cdot |$ is defined to be the number
of vertices plus one),
\( \tilde{\Delta}   = \log \frac{n_1 n_2}{n_1 +n_2}\).
\begin{enumerate}
\item If $p\in A(\D)$, then we construct
$A(\D')=A(\D) \cup \{ p'_1, p'_{2} \} \setminus \{ p \}$. $B(\D')$ is
connected, and $\Delta = \log \frac{n_1 n_{2}}{n_1 +n_{2}}$, so \( \Delta
\geq \tilde{\Delta}. \)

\item If $p \not \in A(\D)$, then there
is a simple path in $B(\D)$ between $u$ and $v$. Without loss of generality,
assume the path passes through $p'_{2}$. Construct $A(\D')=A(\D) \cup \{p'_2
\} \setminus \{p\}$. $B(\D')$ is connected, and $\Delta = \log n_{2}$, so \(
\Delta \geq \log n_{2} + \log \frac{n_1 }{n_1 + n_{2}} = \tilde{\Delta}.\)
\end{enumerate}

\item $u, v$ are on different paths, $u\in p$ and $v\in q$,
for some $p, q \in \PP$ and $p \neq q$. In this case $|\PP'|=|\PP| +2$.
Assume $p=p'_{1} \cup \{u\} \cup p'_{2}$ and $q= p'_{3} \cup \{ v \} \cup
p'_{4}$. Here \( \tilde{\Delta}= \log \frac{n_{1} n_2 n_{3}
n_{4}}{(n_{1}+n_{2})(n_{3}+n_{4})}.\)
\begin{enumerate}
\item If $p, q \in A(\D)$. Without loss of generality  assume that
$n_{1}= \min \{ n_{1}, n_2, n_{3}, n_{4} \}$. Construct $A(\D')=A(\D) \cup \{
p'_2, p'_{3}, p'_{4} \} \setminus \{p, q \}$. Obviously $B(\D')$ is
connected, and $\Delta=\log \frac{n_2 n_{3}
n_{4}}{(n_{1}+n_{2})(n_{3}+n_{4})}$. Without loss of generality, assume
$n_{3} \leq n_{4}$, so \( \log \frac{n_2 n_{4}}{(n_1 + n_2) (n_{3} + n_{4})}
\geq -2\). Thus
\begin{multline*}
\Delta  = \log n_{3} + \log \frac{n_{2}n_{4}}
{(n_{1}+ n_{2})(n_{3} + n_{4})} \geq  \log n_3 -2 \geq \tfrac{1}{2}\log n_3 \geq\\
 \tfrac{1}{4} (\log n_3 + \log n_1 )
 \geq  \tfrac{1}{4} \bigl[  \log n_{3} + \log n_{1}+
 \log \frac{n_{2} n_{4}}
   {(n_{1}+ n_{2})(n_{3} + n_{4})}
\bigr]      = \tfrac{1}{4} \tilde{\Delta}.
\end{multline*}

\item If $p \in A(\D)$, and $q \not \in A(\D)$
(The case $q \in A(\D)$ and $p\notin A(\D)$ is similar). Without loss of
generality, $n_{1} \leq n_2$. $p'_1$ has two endpoints, one of which is
adjacent to $u$.
The other endpoint of $p'_1$ is connected via a simple path in $B(\D)$ to
$v$. This path must contain either $p'_3$ or $p'_4$. Without loss of
generality, assume it contains $p'_{3}$. If $n_{1} \geq n_{3}$ then we
construct $A(\D')=A(\D) \cup \{ p'_{1} , p'_{2} \} \setminus \{ p \}$. If,
on the other hand, $n_{1} < n_{3}$ then we construct $A(\D')=A(\D) \cup \{
p'_{3} , p'_{2} \} \setminus \{ p \}$. In either case, $B(\D')$ is
connected, and
\begin{multline*}
 \Delta = \log \frac{ \max
\{ n_{1}, n_{3} \} \cdot n_2} {n_{1} + n_{2}} =
 \log  \max \{n_{1}, n_{3}\} + \log \frac{n_{2}}{n_{1}+
 n_2}  \geq  \\
 \tfrac{1}{2} \max\{ \log n_1, \log n_3\} \geq \tfrac{1}{4} (\log n_1 +
 \log n_3 ) \geq
 \tfrac{1}{4} (\log n_{1} + \log \min\{n_{3}, n_{4} \} )  \geq\\
   \tfrac{1}{4} \Bigl[ \log n_{1} + \log \min \{n_{3}, n_{4} \} +
   \log \frac{n_2 \max \{n_{3},n_{4}\}}{(n_{1}+ n_{2})(n_{3} + n_{4})}
              \Bigr] = \tfrac{1}{4} \tilde{\Delta}
  \end{multline*}

\item
If $p, q \not \in A(\D)$. Then there is a simple path in $B(\D)$ from $u$ to
$v$. Without loss of generality, assume it passes through $p'_1$, and $p'_3$.
Without loss of generality, assume $n_1 \leq n_3$. Construct $A(\D')=A(\D)
\cup \{ p'_3 \} $. $B(\D')$ is connected, and
\[
 \Delta= \log n_3 \geq \frac{1}{2} \bigl[
   \log \min \{n_1, n_2 \} +  \log \min\{ n_3, n_4 \}
   \bigr] \geq \tfrac{1}{2} \tilde{\Delta}
\]
\end{enumerate}

\item $u, v$ are on the same path. In this case $|\PP'|=|\PP|+ 2$.
Assume $p=p'_{1} \cup \{u\} \cup p'_{2} \cup \{ v \} \cup p'_{3}$, for some
$p \in \PP$ and $p'_1,p'_2, p'_3 \in \PP'$. So $\tilde{\Delta}= \log
\frac{n_1 n_{2} n_{3}} {n_1 + n_{2} + n_3}$. Here $e=u v$ does not
contradict $\D$ being proper, 
implying that $n_{2} \geq (n_1+ n_{2} + n_{3}) ^{1/4}$, so $\log n_{2} \geq
\tfrac{1}{4} \log \max \{ n_1, n_{2} , n_{3} \}$. Without loss of
generality, assume that $n_1 \leq n_{3}$.

\begin{enumerate}
 \item If $p \in A(\D)$, then construct
$A(\D')=A(\D) \cup \{ p'_{2}, p'_{3} \} \setminus \{ p\}$.
\begin{multline*}
\Delta= \log \frac {n_{2} n_{3}} {n_1 + n_{2}+ n_{3}}= \log \min \{ n_2, n_3
\} + \log \frac {\max\{ n_{2}, n_{3} \}} {n_1 + n_{2}+
n_{3}} \geq \\
 \geq \log \min \{ n_2, n_3 \} - \log 3 \geq \tfrac{1}{2} \log \min \{ n_2, n_3
   \} \geq \\
 \geq \tfrac{1}{10} ( \log n_1 + \log \min \{ n_2,n_3 \}) \geq
        \tfrac{1}{10} \tilde{\Delta}
\end{multline*}

\item If $p \not \in A(\D)$, then construct $A(\D')= A(\D) \cup \{ p'_{2} \}$.
\[  \Delta  = \log n_2 \geq
       \tfrac{1}{8}2 \log \max \{ n_1, n_2, n_3 \} \geq
       \tfrac{1}{8}\tilde{\Delta}
\]
\end{enumerate}
\end{enumerate}
\end{proof}

We combine lemmas~\ref{lemma:decompose-proper} and
\ref{lemma:vine-selection}:
\begin{proof}[Proof of Lemma~\ref{lemma:full vine selection}]
We first find an initial decomposition (not necessarily proper) and a
corresponding initial {\gvs} as required by
Lemma~\ref{lemma:vine-selection}. For a proper path,
Lemma~\ref{lemma:decompose-proper} gives the needed initial decomposition and
\gvs. For a complex $(B, \Q)$, we take $(\emptyset, \Q)$ as the initial
decomposition, and $\Q$ as the corresponding initial {\gvs}. Using
Lemma~\ref{lemma:vine-selection} we obtain an irreducible decomposition and
a corresponding \gvs.
\end{proof}

\subsubsection*{Proof of Lemma~\ref{lemma:gvd}}


We need the following definitions:

\begin{definition}
Define a sequence $(c_m)_{m\in \mathbb{N}}$  recursively as follows
 \[ c_m=  \begin{cases}
       c/4 & m<8^4; \\
       c_{\lceil m^{1/4} \rceil } \cdot
       \left(1- \frac{5}{m^{3/16}}\right) & \text{otherwise.}
       \end{cases}
       \]
We define a \emph{\gvd}  $\V$ of a proper path $q$ related to a recursive
decomposition $\D^R(q)$, to be a proper vine decomposition of $q$ such that
\begin{itemize}
\item For every path $p \in \PP$, $|p| < |q|^{1/4}$.
\item $\pval(\V) \geq c_{|q|} \cdot \left [\tval(\D^R(q))-  \log |q| \right ]$.
\end{itemize}
\end{definition}

\begin{proposition}
 There exists $c'>0$ such that for all $m$,    \( c_m \geq c' \).
\end{proposition}
\begin{proof}
Define $c'_j = \frac{c}{4} \prod_{i=2}^{j} (1 - \frac{5}{8^{(4^i)\cdot
3/16}})$, and $c'=\lim_{j \rightarrow \infty} c'_j$. It is easily seen that
$(c_m)_{m>0}$ and $(c'_j)_{j>0}$ are non-increasing and
$c_{8^{4^j}} = c'_j$. We have $c'>0$, since
 $1 > \frac{5}{8^{(4^i)\cdot 3/16}} \geq 0$ and
      $\sum_{i=1}^{\infty} \frac{5}{8^{(4^i)\cdot 3/16}} < \infty$.
\end{proof}

Thus, constructing a recursive irreducible decomposition along
with {\gvd} would suffice to prove Lemma~\ref{lemma:gvd}. The
following lemma is the inductive argument that allows the
construction.

\begin{lemma}
\label{lemma:inductive gvd} Let $\D^R(q)=(S, \{ \D^R(p_1),\dots,\D^R(p_s)
\})$ be a recursive decomposition of a proper path $q$ such that
$\D=(S,\{p_1,\ldots,p_s\})$ is an irreducible decomposition. Assume that:
\begin{itemize}
\item There is a {\gvs} $A(\D)$ of $\D$.
\item For every sub-path $p_i$, there is a
      {\gvd}  $\V(p_i)$ related to $\D^R(p_i)$.
\end{itemize}
Then we can construct $\V$, a {\gvd}  related to $\D^R(q)$.
\end{lemma}
\begin{proof}
Denote by $m_i=|p_i|$ and
$M=|q|$. From the irreducibility property of $\D^R(q)$, $m_i \leq
M^{{1}/{4}}$. For every $p\in A(\D)$  we do the following:

\newcommand{\Sl}{C^{\text{lft}}}
\newcommand{\Sm}{C^{\text{mid}}}
\newcommand{\Sr}{C^{\text{rgt}}}
\newcommand{\Pl}{\ensuremath{p^{\text{lft}}}}
\newcommand{\Pm}{\ensuremath{p^{\text{mid}}}}
\renewcommand{\Pr}{\ensuremath{p^{\text{rgt}}}}
\newcommand{\Vl}{\ensuremath{\V^{\text{lft}}}}
\newcommand{\Vr}{\ensuremath{\V^{\text{rgt}}}}
\newcommand{\Bl}{\ensuremath{B^{\text{lft}}}}

\begin{trivlist}
\item
Let $m=|p|$.  From the assumptions, the vines in $\V(p)=(B(p),\PP(p))$ have
lengths at most $m^{1/4}$. We partition $p$ to $t
\defd \lceil m/(5 m^{1/4}) \rceil= \lceil m^{3/4}/5 \rceil $ consecutive
sub-paths $C_1,\dots,C_t$, each of size between $(5 - 2) m ^{1/4}$ and $(5 +
2) m^{1/4}$, such that no sub-path partitions a vine in $\V(p)$. Let $X_i$ be
the sum of the values of the vines in $C_i$. Thus $\sum _i X_i=
\pval(\V(p))$. Let $C \defd S_{i_1}$ be the sub-path that satisfies $X \defd
X_{i_1}=\min _i \{X_i\}$, so $X \leq \frac{5}{m^{3/4}} \pval(\V(p))$. $S$ is
partitioned into three consecutive sub-paths $\Sl, \Sm, \Sr$ such that
$|\Sm|=m^{1/4}$, and $|\Sl|\geq m^{1/4}$, $|\Sr|\geq m^{1/4}$.

Let $\Pl$, $\Pr$ denotes  the sub-paths $p$ that are to the left and to the
right (respectively) of $\Sm$ in $p$. Let $B(p)$ denotes the backbone of
$\V(p)$. We construct $\Vl$, a vine decomposition of $\Pl$
by defining
$\Bl$, the backbone of $\Vl$, to be $\Bl= (B(p)\cup \Sl) \cap \Pl$. We claim
that $\Vl$ is a legal vine decomposition, and that the endpoints of $\Pl$ are
part of $\Bl$. It is easy to see  that the endpoints of the vines in $\Vl$
are adjacent to $\Bl$ and the endpoints of $\Pl$ are part of $\Bl$.

We are left to prove that the sub-graph spanned by $\Bl$ is connected. First,
note that $\exists v_l \in \Sl \cap B(p)$, otherwise $\Sl$ is part of a vine
in $\V(p)$ and its length is at least $m^{1/4}$, which contradicts the
assumptions. The sub-graph induced on $\Sl$ is obviously connected. For every
vertex $u\in \Bl \setminus \Sl$, there is a {path} $r$ in $B(p)$ from $u$ to
$v_l$. Consider the maximal prefix of $r$ which is entirely inside $\Pl$.
The last vertex of the maximal prefix of $r$ must be inside $\Sl$, otherwise
there is an edge $e$ in $p$ such that $|e|_{p} \geq |\Sl|=m^{1/4}$, in
contradiction to the irreducibility property of $\D$. Thus $\Bl$ is
connected.

In conclusion, $\Vl$ is a legal vine decomposition of $\Pl$. Similarly, we
construct a vine decomposition $\Vr$ of $\Pr$ such that the endpoints of
$\Pr$ are part of the backbone of $\Vr$.
\end{trivlist}

We construct $\V$, a vine decomposition of $q$ such that every vertex in a
sub-path $p \not \in A(\D)$ will have the same role (part of the backbone or
part of a vine) as in $\V(p)$. For $p \in A(\D)$ we declare $\Sm(p)$ to be a
vine, and the other vertices, to have the same role as in $\Vl(p)$ and
$\Vr(p)$. $\V$ is a vine decomposition of $q$ as it is seen from the
previous discussion and because the endpoints of the sub paths are part of
the backbone (see Def.~\ref{def:vine proper}). Also, every vine of $\V$ has
length less than $M^{1/4}$.

Because $c_{m'}$ is non-increasing as a function of $m'$, $c_M=
\min_{m' \leq M^{1/4}} c_{m'} (1-\frac{5}{m'{}^{3/4}})$. Let
$m=|p|$. From the construction, for every $p\in A(\D)$,
\begin{multline*}
\pval(\Vl(p)) + \pval(\Vr(p)) \geq \left (1-\frac{5}{m^{3/4}} \right )
       \pval(\V(p)) \geq \\
       \geq  \left ( 1-\frac{5}{m^{3/4}} \right )
        c_m \left (\tval(\D^R(p))- \log m \right )
\geq c_M \left (\tval(\D^R(p))- \log m \right ).
\end{multline*}
We conclude
\begin{align*}
\pval(\V) & \geq \sum_{p \in A(\D)} \tfrac{1}{4} \log |p|  +
          \sum_{p\in A(\D)} c_M \left[\tval(\D^R(p)) -
         \log |p| \right]  +\\
 &    \quad  + \sum_{p \not \in A(\D)} \left[\tval(\D^R(p)) -
         \log |p| \right]  \\
    & \geq c_M \cdot \left [\sum_{i=1}^s \log |p_i| +
     \sum_{i=1}^s \left [\tval(\D^R(p_i)) -  \log |p_i| \right ] \right ] \\
&   = c_M \sum_{i=1}^s \tval(\D^R(p_i))
    = c_M \left [ \tval(\D^R(q)) -\log M \right ]. \qed
\end{align*}
\renewcommand{\qed}{}
\end{proof}

\begin{proof}[Proof of Lemma \ref{lemma:gvd}]
We show the existence of a recursive decomposition and
{\gvd} by induction on the structure of the recursive
decomposition.

The base of the induction are simple paths whose inner vertices do not have
other adjacent vertices. Such a path has a unique recursive decomposition,
namely all the vertices are in the separating set of the decomposition. The
{\gvd} is also obvious: all the vertices are in the backbone.

Otherwise, we use Lemma~\ref{lemma:full vine selection} to get an irreducible
decomposition $\D=(S,\PP)$ and a vine selection $A(\D)$. Inductively, Each
$p\in \PP$ has a recursive decomposition
$\D^R(p)$ along with a {\gvd} $\V(p)$.
Hence, all the assumptions of Lemma~\ref{lemma:inductive gvd} are
satisfied, and we get a recursive decomposition along
with a {\gvd} for the given proper path.
\end{proof}

\subsubsection*{Proof of Lemma~\ref{lem:throw-h}}

\begin{proof}[Proof of {Lemma~\ref{lem:throw-h}}.]
Assume that the first two cases in the statement of the Lemma do not hold.
Denote $\Q= \{q_1,\dots,q_t\}$. We divide the vertices of $q_1 \cup \dots
\cup q_t$ between four sets $X_1$, $X_2$, $X_3$ and $X_4$ as follows: Let
$v_1,v_2,\dots, v_m$ be the vertices of $q_1\cup q_2 \cup \dots \cup q_t$
ordered as the order in $q_1,q_2,\dots,q_t$. Let $t_0=1$ and $t_4=m$. For
$i\in \{1,2,3\}$, let $t_i$ be the smallest index such that $t_i \geq
t_{i-1}+h+1$ and $v_{t_i}$ is not on a path of $\PP$. Now we assign
$X_i=\{v_{t_{i-1}},\dots, v_{t_i} \}$. Note that the vertices
$v_{t_1},v_{t_2},v_{t_3}$ might appear in two consecutive sets.

All four sets have at least $h$ unique vertices, unless there exists a path
$p\in \PP$ such that $|p| \geq h+1$. In this case we construct $\V'$ from
$\V$ by removing the path  $p$. Here $\vval(\V')=  \vval(\V)-\log |p| \geq
\tfrac{23}{24}\vval(\V))$.

 From here on, we assume all four sets has at least $h$ unique vertices. We
denote by $\vval(X_i)=\sum_p \log |p|$ where $p$ ranges over the paths in
$\PP$ contained in $X_i$. Thus $\sum _{i=1}^4 \vval(X_i) = \vval (\V) $.
Assume that $\vval(X_{i_4})=\max _{i\leq 4} \vval(X_i)$, so $\vval(X_{i_4})
\geq \vval(\V) / 4 $.

For a vertex $u\in X_{i_4} \cap B_2$, define
 \[ C_u= \Bigl \{v \in (q_1\cup\cdots \cup q_t)\setminus X_{i_4}: \
  \begin{aligned}
  \exists r=\la u,w_1,\ldots,w_s,v \ra \text{ a
 path in } G^s  \\
 \text{ and } \{w_1,\ldots,w_s\}\subseteq X_{i_4}\cap B_2
 \end{aligned}
 \Bigr \}.
 \]
Note that $C_u \neq \emptyset$. We say that $u\in X_{i_4} \cap B_2$ is
\emph{locally connected} to $A$ if $C_u \subseteq A$. Let $Y_{i_j} (j \leq
3)$ be the set of paths of $\PP$ contained in $X_{i_4}$ having at least one
of their endpoints locally connected to $X_{i_j}$. Then
$\vval(Y_{i_1})+\vval(Y_{i_2})+\vval(Y_{i_3}) \leq 2 \vval(X_{i_4})$, since
each path of $\PP$ contained in $X_{i_4}$ is counted at most twice. Assume
that $Y_{i_1}$ has the minimal value of the three, so $\vval(Y_{i_1}) \leq
\frac{2}{3} \vval(X_{i_4})$.

We remove from $\V$ all the unique vertices of $X_{i_1}$. Note that there are
indeed enough vertices to remove since $X_{i_1}$ alone contains at least $h$
vertices that can be removed.

Next, We add to the backbone the following vertices: (i) The vertices of the
vines in $Y_{i_1}$;  and (iii) {\em all} the vertices in $X_{i_2}$ and
$X_{i_3}$ to the backbone. The resulting is a ``vine decomposition''
$\tilde{\V}=(\tilde{B},\tilde{\PP})$ that the endpoints of its vines are all
part of its backbone, and $\vval(\tilde{\V}) \geq \frac{1}{3} \vval(X_{i_4})
\geq \frac{1}{12} \vval(\V)$.

However, $\tilde{B}$ might be disconnected. It consists of at most 3
connected components: (i) One that contains  $B_1$, (ii)  one that contains
$X_{i_2}$, and (iii) one that contains $X_{i_3}$. A more careful observation
reveals that there might be at most 2 connected components: Assume $X_{i_2}$
is disconnected from $B_1$ in $\tilde{B}$. In this case $X_{i_2}$ is
contained in one path $q_{i_0}\in \Q$. On one side of $X_{i_2}$ (in
$q_{i_0}$), there are vertices from $X_{i_1}$, and in the other side,
vertices from $X_{i_4}$. Now, either $X_{i_3}$ contains an endpoint of one
of the paths of $\Q$ (in this case it is connected to $B_1$) or it is
adjacent to $X_{i_2}$ in $q_{i_0}$ (in this case it is connected to
$X_{i_2}$). In both cases we have only two connected components.

In case $\tilde{B}$ is disconnected we augment it as follows. Denote the
connected components of $\tilde{B}$ by $C \supseteq X_{i_2}$ and $D \supseteq
B_1$. On $q_{i_0}$, $X_{i_4}\cap q_{i_0}$ separates $X_{i_2}$ from $D \cap
(q_{i_0} \cup \{q_{i_0}$'s neighbors in $B\})$.

Assume that $C$ is on the left side of $X_{i_4}\cap q_{i_0}$ in $q_{i_0}$.
Let $u\in C\cap q_{i_0}$ be the right most vertex of $C$ in $q_{i_0}$. $u$
is either in $X_{i_4}$ or adjacent to $X_{i_4}$. Let $v \in D$ be the
leftmost vertex of $D$ that is to the right of $u$ (in case no vertex of
$q_{i_0}$ right of $u$ is in $D$, we take $v$ to be a vertex in $B_1$
adjacent to the rightmost vertex in $q_{i_0}$). The vertices of $q_{i_0}$
between $u$ and $v$ form a path $p\in\PP$ which is also a vine in
$\tilde{\V}$. We add them to the backbone of $\tilde{\V}$. The result is the
desired vine decomposition $\V'$, which has at least $h$ vertices less than
$\V$, and
 \[ \vval(\V')\geq  \vval(\tilde{\V}) - \log |p|  \geq
 \tfrac{1}{12} \vval({\V}) - \tfrac{1}{24}\vval({\V}) \geq \vval(\V)/24. \]

We observe that from the way $X_i$ were formed, for any path $q\in \Q$,
$q\cap X_{i_4}$ is indeed a subpath, and one of its endpoints is adjacent to
the backbone of $\V'$.
\end{proof}

\section{Refined Locality of Reference} \seclab{extended}

\begin{definition}
An \emph{extended access graph} on a set of pages $\mathcal{P}$  is a finite
labeled undirected graph $G=(V,E,\ell)$, with a labeling function
$\ell:V\rightarrow {\cal P}$ that labels vertices with pages.

A request sequence $\sigma=(\ell(v_i))_{i \geq 1}$, is a finite sequence of
labels attached to vertices from $G$, such that either $v_{i+1}=v_{i}$, or
$v_i v_{i+1} \in E$. A paging algorithm should maintain the invariant, that
following the $i$th request, $\ell(v_i)$ should be in some page slot. The
competitive measures $\compr_{\on}(G,k)$, $\compr(G,k)$, $\compr_{\obl}
(G,k)$ and their asymptotic counterparts are defined in a similar manner to
the definitions for access graphs in \secref{locality}.
\end{definition}

For a given extended access graph $G$, we define a parameter $\Delta(G)$
that indicates ``how quickly" the locality of reference may change $G$.

\begin{definition}
\[ \Delta(G)=\min \{ s-1 :\; \la v_1, v_2,\dots,v_s \ra \in \text{paths}(G),\
   v_1 \neq v_s \text{, and } \ell(v_1)=\ell(v_s) \}.
\]
As convention, when $G$ is an (non extended) access graph, we fix
$\Delta(G)=\infty$.
\end{definition}

We remark that paging algorithms get only the names of the requested pages
and not the names of the vertices. However for $\Delta(G)>2$ and assuming
the starting vertex is known, online algorithms can easily reconstruct the
sequence of the requested vertices from the sequence of the requested pages.

\subsection{Truly Online Algorithms} \seclab{eag:to}

\theoref{rto} and \theoref{dto} extend to the extended access graph model.

\begin{theorem}
For any $\e>0$, and any extended access graph $G$ with $\Delta(G) > (1+\e)k$,
\begin{align*}
 \compr_{\text{\ddet}}(G,k) & \leq \max \{O(\compr^\infty(G,k)), 2/ \e \} \\
 \compr_{\text{\drand}}(G,k)& \leq \max \{ O(\compr^\infty_{\obl}(G,k)), 2/ \e \}.
\end{align*}
In particular, for any fixed $\alpha>1$, {\ddet} and {\drand} are very
strongly competitive for any $G$ and $k$ satisfying $\Delta(G)>\alpha k$.
\end{theorem}
\begin{proof}
Let $g$ denote the number of new pages in the current phase. We consider two
cases: If $g \geq \e k$ then  {\ddet} and {\drand} fault at most $k \leq
\frac{2}{\e} \cdot \frac{\e k}{2}\leq \frac{2}{\e} \cdot \frac{g}{2}$ times
in the phase.

If, on the other hand, $g< \e k$, then during the current phase and the
previous phase there were no requests to two different vertices with the
same label, so the graph $\GIII$ is an actual sub-graph of $G$. By the
proofs of \theoref{dto} and \theoref{rto}, {\ddet} faults at most $O(g \cdot
\compr^\infty(G,k))$ times and {\drand} faults at most $O(g \cdot
\compr_{\obl}^\infty(G,k))$ times, in the phase.
\end{proof}

When $\Delta(G)$ is slightly greater than $k$, {\ddet} and {\drand} do not
work well, as is seen in the following example.
\begin{example}
\label{exam:2} Consider the extended access graph $G=(V,E,\ell)$, where
$V=\{1,2,\ldots,v_{k+1},v_{k+2}\}$, $E=\{v_i v_{i+1}:\; 1\leq i \leq k+1\}$,
and
 \[ \ell(v_i)= \begin{cases} i & i \leq k+1\\ 1 & i=k+2 .\end{cases}  \]
Here $\Delta(G)=k+1$. Consider the request sequence $\sigma=J_1 J_2
J_3\ldots$ where $J_{2i-1}=1,2,\ldots,k$ and $J_{2i}=k+1,1,k+1,k,\ldots,2$.
Note that $(J_i)_i$ is the phase partitioning of $\sigma$. Clearly,
$r(G,k)=O(1)$. In contrast, {\ddet} and {\drand} fault $\Omega(\log k)$
times [expected] during subphase III in each phase, and therefore
$r^\infty_{\dto}(G,k)$ and $r^\infty_{\rto}(G,k)$ are $\Omega(\log k)$.
\end{example}

\subsection{Very strongly competitive algorithm for paths with $k+1$ pages}

In this section we present a very strongly competitive algorithm for graphs
that are simple paths on $k+1$ pages. This algorithm will serve us in proving
impossibility results concerning truly online algorithms.

Given a finite simple path $G=\la v_1,v_2,\ldots,v_m \ra$ with
a surjective label function $\ell:\{v_i:\; 1\leq i\leq
m\}\rightarrow {\cal P}$, where $|{\cal P}|=k+1$.
Fix a vertex $v_i$. We define a partial order $\prec_i$ on $\cal
P$: $p \prec_i q$ if any path from $v_i$ to a vertex labeled by
$q$ must include a vertex labeled by $p$. It is easy to verify
that $\prec_i$ is indeed a partial order. Let $M_i$ be the set of
maximal elements in $\prec_i$. Note that:
\begin{enumerate}
\item $M_i$ includes at most two pages that appear only on one side of
$v_i$ in $G$.
\item Any request sequence that starts at $v_i$ and accesses all the pages in
$M_i$, must access all the pages in $\cal P$.
\end{enumerate}

We define linear orders on $M'_i=\{p \in M_i:\; p$ appears on both sides of
$v_i$ in $G\}$. For $p,q\in M'_i$, $p \prec^L_i q$ if when going in $G$ from
$v_i$ to the left, we reach a vertex labeled with $p$ before we reach  a
vertex labeled with $q$. Analogously, $p\prec^R_i q$ is defined to the
right. Note that $p \prec_i^R q$ if and only if  $q \prec_i^L p$.

\begin{lemma}
$\compr^\infty_\obl (G,k)= \Omega (\max_i \log |M_i|)$.
\end{lemma}
\begin{proof}
Let $i_0= \arg \max_i |M_i|$,  so $|M_{i_0}|>0$. Using Yao's Principle
(cf.~\cite[Ch.~8]{BEY98}) we construct a probability distribution on the
request sequences by the following iterative process: The request sequence
is composed of \emph{periods}, each period is composed of \emph{sub-periods}.
Let $N_j$ be the set of $\prec_{i_0}$--maximal pages that were left unmarked
before the $j$th sub-period begins. At the beginning of a period $N_1=M_{i_0}$.

We describe the request sequence in the $j$th sub-period. The adversary
begins from $v_{i_0}$. It chooses a page $p\in N_j$ such that $p$ is smaller
than $|N_j|/2 + 1$ pages of $N_j$ under both $\prec^L_{i_0}$ and
$\prec^R_{i_0}$ (its existence follows from a simple counting argument). If
$p$ appears on both sides of $v_{i_0}$ in $G$, then the adversary chooses
uniformly at random the left or the right side, otherwise he chooses the side
on which $p$ appears. The adversary then requests the vertices in that
direction until reaching the first vertex $v$ labeled with $p$, and then
returns to $v_{i_0}$. At this point the $j$th sub-period ends. Note that
$|N_{j+1}| \geq |N_j|/2 -1$. The adversary continues this way until
$N_j=\emptyset$, which means that all pages were requested during the period.
At this point the period ends, and the adversary begins a new period.

During a period, an optimal off-line algorithm faults at most twice, because
the request sequence consists of at most two phases.

Next we show that any online algorithm has $\Omega(\log |M_{i_0}|)$ expected
number of faults of during a period.

To prove it we argue that there are $\Omega(\log|M_{i_0}|)$ sub-periods in a
period and an online algorithm has an expected cost of at least half in each
sub-period, except maybe two of them. There are at least $\Omega(\log
|M_{i_0}|)$ sub-periods before  $N_j$ becomes empty set, because in each
sub-period the size of $N_j$ is roughly halved. In all sub-periods in which
the target label $p$ appears on both sides of $v_{i_0}$, the expected number of faults of the
online algorithm is at least $1/2$, since $\prec_{i_0}$-maximal pages that
appear on both sides of $v_{i_0}$ must appear, at least on one side of
$v_{i_0}$, after the hole. As all $\prec_{i_0}$-maximal pages appear on both sides
of $v_{i_0}$, except maybe two, the claim follows.
\end{proof}

We shall see now a very strongly competitive deterministic marking online
algorithm called {\maxfar}.

\paragraph{{\small MAXFAR}.} {\maxfar} is a marking algorithm.
Assume the current phase began at $v_{i}$. On the $j$th fault in the phase,
let $N_j\subset M_{i}$ be the set of unmarked $\prec_{i}$--maximal elements.
Choose a page $p\in N_j$ in the middle of $N_j$ according to $\prec_i^L$
(which is also in the middle according to $\prec_i^R$) and evict it. If
$N_j$ contains only pages that appear on only one side of $v_i$, evict one
of them (there at most two such pages).
\medskip

It is easy to see that $\{N_j\}_j$ is a decreasing sequence of sets, and
$|N_{j+1}|\leq |N_j|/2 +1$. Thus after at most $j=O(\log |M_{i_0}|)$ faults,
$N_j= \emptyset$, and at this point, the phase is over. We conclude that
$\compr_{\text{\maxfar}}(G,k)= O( \max _i \log |M_{i}|)$. Putting this
together,

\begin{theorem} \theolab{maxfar}
For any extended access graph $G$ which is a simple path on $k+1$ pages, \(
\compr_{\text{\maxfar}}(G,k)=O(\compr^\infty_\obl(G,k)) . \)
\end{theorem}

\subsection{Impossibility Results} \seclab{eag:impossibility}

In this section we show that any {\em truly online} algorithm can not be
very strongly competitive on extended access graphs, when $\Delta(G)$ is
slightly less than $k$. Formally, we prove that

\begin{theorem}
\begin{enumerate}
\item
For any $0<f<k$, and any deterministic truly on-line paging algorithm $A$,
there exists an extended access graph $G$ such that $\Delta(G) \geq k - f+1$
and $\compr_{A}(G,k)\geq f$ but $\compr(G,k)=O(\log k)$. In particular, for
$f=f(k)=\omega(\log k)$, $r_A(G,k)=\omega(r(G,k))$.
\item
For any $0<f<k$, and any randomized truly on-line paging algorithm $A$,
there exists an extended access graph $G$ such that $\Delta(G) \geq k- f-1$
and $\compr_{A}(G,k)= \Omega(\log f)$, but $\compr_{\obl}(G,k)=O(\log k-
\log f +\log \log f)$. In particular, for $f=f(k)=k^{1-o(1)}$,
$\compr_{A}(G,k)=\omega(\compr_{\obl}(G,k))$.
\end{enumerate}
\end{theorem}
\begin{proof}
Fix $0<f'=f+1<k$, and a truly online algorithm $A$.

Consider the following {\sl part} of an extended access graph
labeled with $k+1$ pages:
\[ 1,2,\ldots,k,k+1,x_1,x_2,\ldots,x_{f'} .\]
The adversary maintains the invariant $\{x_i:\; 1\leq i \leq
f'\}=\{1,\ldots,f'\}$, but in some permutation that will be determined.
\medskip

To prove part (1), assume $A$ is deterministic. We split the request
sequence $\sigma$ into phases and show that each phase costs $A$ at least
$f'$, whereas it costs the adversary only $1$ (except the first phase, in
which the cost for the adversary is $k$). We will also show that there exists
an online algorithm that obtains a competitive ratio of $O(\log k)$ on this
graph. By making the request sequence long enough (for $c$ phases where $c$
satisfies $k+(c-1)f'\geq fc$),  the first part of the theorem will be proved.

The adversary works in phases. Assume that in the previous phase the
requested pages were $\{1,\dots,k\}$. Each phase is composed of $f'$
sub-phases, indexed by $j$

For $j=0$, the adversary requests $x_0=k+1$. As $A$ is deterministic, the
adversary knows what page  has been evicted by $A$ to satisfy this request.
Denote it by $p_0$. In general, in each subphase, the adversary requests a
hole of $A$, and therefore to serve the request, $A$ must evict at least one
page from its real memory. Denote a page evicted by $A$ in the $j$th
sub-phase by $p_j$.

In the $j$th sub-phase the adversary does as follows: If $p_{j-1}>f'$ then
the adversary requests the pages $x_{j-1}, x_{j-2}, \ldots, x_0=k+1, k,
\ldots, p_{j-1}+1, p_{j-1}, p_{j-1}+1, p_{j-1}+2, \ldots, k, x_0 = k+1, x_1,
\ldots, x_{j-1}.$ It then sets $x_j$ to be arbitrary page in
$\{1,\ldots,f'\}\setminus\{x_1,\ldots,x_{j-1}\}$ and requests $x_j$. This is
a legitimate traversal on the extended access graph.

If $p_{j-1}\leq f'$ then: (i) If $p_{j-1}$ has not been requested yet in the
current phase, then the adversary sets $x_j=p_{j-1}$, and requests it (ii)
Otherwise there must be some $i\leq j$ such that $x_i=p_{j-1}$ and the
adversary generates the requests $x_{j-1}, x_{j-2}, \ldots, x_i, x_{i+1},
\ldots, x_{j-1}$. It then sets $x_j$ to be arbitrary page in
$\{1,\ldots,f'\}\setminus\{x_1,\ldots,x_{j-1}\}$ and requests $x_j$.

When $j=f'$,  the  phase ends. $k$ different pages have been requested in
this phase. $A$ had a fault in each sub-phase and therefore its cost has
been at least $f'$.

The above argument is for one phase, but we can continue this process for an
additional $c$ phases, for any $c$, simply by adding an additional $c\cdot
f'$ vertices on the right hand side (right of $x_{f'}$ above).

Note that (i) $\Delta(G)\geq k - f'$ for any such graph; (ii) $G$ is a
simple path on $k+1$ pages, so from \theoref{maxfar}, {\maxfar} has a
competitive ratio of $O(\log k)$.

\bigskip
Next, we turn to prove an impossibility result for randomized truly online
algorithms. The adversary constructs a graph similarly to the previous case,
but now it does not know where the hole is. Instead it resorts to a random
process as follows: It maintains a maximal unrequested sub-path $\la
i_1,\dots, i_2 \ra$ of the vertices labeled $1,\dots, f'$. In each sub-phase
it first requests all the pages already requested during the phase. The
adversary then sets $i_3=(i_1+i_2)/2$, the midpoint vertex in the unrequested
segment, and chooses uniformly at random one of the two:
\begin{itemize}
\item Requesting the pages $\{i_3,\dots,i_2\}$ by setting
$x_{j+1}=i_2$, $x_{j+2}=i_2-1$, $\dots, x_{j+(i_2-i_3)+1}=i_3$,
and then setting $j\leftarrow j+(i_2-i_3)+1$.

\item Requesting the pages $\{i_1,\dots,i_3\}$ by setting
$x_{j+1}=i_3$, $x_{j+2}=i_3-1$, $\dots, x_{j+(i_3-i_1)+1}=i_1$,
and then setting $j\leftarrow j+(i_3-i_1)+1$.
\end{itemize}
After approximately $\log f'$ sub-phases the phase ends. In each sub-phase,
$A$ faults with probability at least $1/2$, and therefore the expected cost
of $A$ in a phase is  $\Omega(\log f)$.

Like the deterministic case, this process can be continued for an additional
$c$ phases, for any $c$, simply by adding an additional vertices on the
right hand side.

Note also that the construction maintains $\Delta(G)\geq k- f$,
and that $G$ is a simple path on $k+1$ pages. Thus {\maxfar} is
applicable in this scenario.

Assume $i\geq k+1$. To bound $M_i$, denote by $t$ the first time $v_i$ is
reached by the adversary. Each sub-phase of the adversary after time $t$,
contributes at most one maximal element to $\prec_i$, and thus, at most
$\lceil \log f'\rceil$ maximal elements are added to $M_i$ in each phase
after time $t$. However, after $\lceil (k+1) / f' \rceil$ phases  since time
$t$, all the pages also appear to the right of $v_i$, since each phase adds
$f'$ distinct pages to the right of $v_i$. Therefore $|M_i| \leq \left\lceil
k/f' \right\rceil \left\lceil \log f' \right\rceil$. Note that a similar
argument also holds for $i\leq k$. Thus, $r_{\maxfar}(G,k)= O(\log k - \log
f' +\log \log f')$.
\end{proof}

\section{Implementation} \seclab{truly:implementation}

Storing $G_0$ requires only $O(k \log n)$ bits. We also need to keep track
of:
\begin{enumerate}
\item The vertices requested so far in the current phase.
\item The current unevicted vertices of $G_0$.
\item $G_0$ for the next phase.
\end{enumerate}

One way to construct $G_0$ for the {\em next} phase is to store a
pointer from every page requested to the previously requested
page. This pointer will be updated only once in a phase (the first
time the page is accessed during the phase). The resulting data
structure is a tree with pointers from the leaves
upwards, and the root is the first page requested in the phase.%
\footnote{An alternative approach is to store for each page a
pointer to the {\em next} page, and update it each time the page
is accessed. We get a tree rooted with the last requested page in
the phase. This tree seems to capture better the locality of
reference, as it stores edges resulting from more recent requests.
However, it also has more pointer update operations.} At the end
of a phase, the pointers are scanned and an image of $G_0$ is
built in memory. This image contains vertices, pointers, and the
virtual addresses associated with the vertices. In total (hardware
registers and memory) the memory required is $O(k \log n)$ bits.

There is a rather efficient hardware implementation for processing
page hits. The $\log n$ bit virtual address is translated to a
$\log k$ page slot address (by the virtual address translation
mechanism). The spanning tree is built on the $\log k$ bit address
space of page slots but the virtual addresses ($\log n$ bits) are
stored in the vertices. Hence, the hardware storage requirement
for implementing page hits in {\ddet} and {\drand} is one $\log k$
bits pointer and one extra bit (marking bit) associated with every
page slot. The extra hardware processing associated with
interpreting a page hit is one $\log k$ bits address comparison,
one bit comparison, and (possibly) setting one $\log k$ bits
pointer.

{\ddet} and {\drand}, as presented  here, use a spanning tree of $G_P$ as
their reference access graph. It is easy to check that instead, we could
have use any connected sub-graph of $G_P$. We can even use $G_P$ itself and
store it using only $O(k \log n)$ bits (instead of the $O(k^2 \log n)$ in
the naive implementation). To do that, observe that there is no need to
actually store edges whom both their endpoints have degree $\geq 3$. Thus,
when a new edge is revealed we increase the degree of its
endpoints\footnote{The degree can be stored in only three states:
``degree=1", ``degree=2", and ``degree$\geq 3$".}, and ``forget" all the
edges in which  both endpoints have degree $\geq 3$. In this way only $2k$
edges should be explicitly stored.


It would be of theoretical interest to reduce the memory
requirements even further, to be independent of $n$. Our goal is
to reduce the factor of $\log n$ in the storage requirement to
$\log k$. Of course, the address translation table must still use
$\Omega(k\log n)$ bits. But when we restrict ourselves to the
paging eviction strategy, and assuming we are also told reliably
whether a page request is a hit or a fault, we can allow  small
errors, if their total effect on the number of page faults is
insignificant.

Consider a universal set of hash functions $h:\mathbb{N}
\rightarrow \{0,\dots, m-1\}$, {\sl e.g., } $h(x)= ax+b \pmod{m}$,
where $m$ is prime and $a$ and $b$ are uniformly and independently
sampled from $\ZZ_m$. We replace in {\rto} the usage of the
virtual address of a page $p$ with  the hash value of the page
($h(p)$).

In this case {\rto} may err and identify two different pages as the same. In the
worst case, it would ``confuse" {\rto} in the current and the next phase
(because $G_0$ for the next phase is built incorrectly). Nonetheless, since
{\rto} has the marking property, the damage is restricted
to these two phases, and so such an error  would add at most $2k$ page
faults.

The probability that such a bad event happens is bounded from above by the
following simple argument: consider the $\binom{k+g}{2}$ pairs of different
pages requested during the last two phases. The probability that a pair of
different pages collide is $1/m$, therefore the probability of an error
occurring during a phase is at most $ \binom{2k}{2} \frac{1}{m} =
O\bigl(\frac{k^2}{m}\bigr) $. Choosing $m=\Theta(k^3)$ insures that the
expected number of added page faults due to collisions of hash values is $O(k
\frac{k^2}{m})=O(1)$ per phase.

Hence, a data structure with $O(k \log m)=O(k \log k)$ bits gives
a randomized algorithm with the same performance guarantees as the
original {\rto} algorithm, up to a constant factor.

\section{Concluding Remarks}

In this paper we have studied the access graph model for locality of
reference in paging. We have shown a somewhat surprising result: It is
possible to be both truly online and still very strongly competitive in the
access graph model. The resulting algorithms seems practical, and can be
proven to work well even when the locality of reference changes.

The following issues are not resolved.
 \begin{enumerate}
\item The proof of \lemref{bounded-f3} is very lengthy. Is there a simpler
and/or shorter proof?
\item In a preliminary version of this work \cite{FM97} we claimed that both {\ddet}
and {\drand} can be patched so as to be very strongly competitive for any
$G$ and $k$ as long as $\Delta(G)>k$. The patch we devised turned out to be
erroneous. Still, we conjecture that very strongly competitive truly online
algorithms are possible when $\Delta(G)>k$.
\item Finding an uniform and very strongly competitive algorithm
for the extended access graph model.
\item Is there a very strongly competitive algorithm for the \emph{directed access graph
model}? Or is it computationally a hard problem?
 \end{enumerate}

\paragraph{Acknowledgement.}
We would like to thank Yair Bartal, Nati Linial, and Gideon Stupp for
helpful discussions.

\bibliographystyle{plain}
\bibliography{tpaging}

\begin{thebibliography}{10}

\bibitem{Belady66}
L.A. Belady.
\newblock A study of replacement algorithms for virtual storage computers.
\newblock {\em IBM Systems Journal}, 5:78--101, 1966.

\bibitem{BBKTW94}
Shai Ben-David, Allan Borodin, Richard Karp, G{\'a}bar Tardos, and Avi
  Wigderson.
\newblock On the power of randomization in on-line algorithms.
\newblock {\em Algorithmica}, 11(1):2--14, January 1994.

\bibitem{BEY98}
Allan Borodin and Ran El-Yaniv.
\newblock {\em Online computation and competitive analysis}.
\newblock Cambridge University Press, Cambridge, UK, 1998.

\bibitem{BIRS95}
Allan Borodin, Sandy Irani, Prabhakar Raghavan, and Baruch Schieber.
\newblock Competitive paging with locality of reference.
\newblock {\em J. Comp. Syst. Sci.}, 50(2):244--258, April 1995.

\bibitem{CN99}
Marek Chrobak and John Noga.
\newblock {LRU} is better than {FIFO}.
\newblock {\em Algorithmica}, 23:180--185, 1999.

\bibitem{FK95}
Amos Fiat and Anna~R. Karlin.
\newblock Randomized and multipointer paging with locality of reference.
\newblock In {\em Proceedings of the 27th Annual ACM Symposium on Theory of
  Computing}, pages 626--634, 1995.

\bibitem{FKLMSY91}
Amos Fiat, Richard Karp, M.~Luby, Lyle~A. McGeoch, Daniel~D. Sleator, and
  Neal~E. Young.
\newblock Competitive paging algorithms.
\newblock {\em Journal of Algorithms}, 12:685--699, 1991.

\bibitem{FM97}
Amos Fiat and Manor Mendel.
\newblock Truly online paging with locality of reference (extended abstract).
\newblock In {\em Proceedings of the 38th Annual Symposium on Foundations of
  Computer Science}, pages 326--335, 1997.

\bibitem{FR97}
Amos Fiat and Ziv Rosen.
\newblock Experimental studies of access graph based heuristics: Beating the
  lru standard?
\newblock In {\em 8th Annual ACM-SIAM Symposium on Discrete Algorithms}, pages
  63--72, 1997.

\bibitem{IKP96}
Sandy Irani, Anna~R. Karlin, and Steven~J. Phillips.
\newblock Strongly competitive algorithms for paging with locality of
  reference.
\newblock {\em SIAM J. Comput.}, 25(3):477--497, June 1996.

\bibitem{KPR00}
Anna~R. Karlin, Steven~J. Phillips, and Prabhakar Raghavan.
\newblock Markov paging.
\newblock {\em SIAM J. Comput.}, 30(3):906--922, 2000.

\bibitem{KW91}
Daniel~J. Kleitman and Douglass~B. West.
\newblock Spanning trees with many leaves.
\newblock {\em SIAM Journal on Discrete Mathematics}, 4(1):99--106, 1991.

\bibitem{LPR99}
Carsten Lund, Steven Phillips, and Nick Reingold.
\newblock Paging against a distribution and ip networking.
\newblock {\em J. Comp. Syst. Sci.}, 58(1):222--232, 1999.

\bibitem{SleTar85a}
Daniel~D. Sleator and Robert~E. Tarjan.
\newblock Amortized efficiency of list update and paging rules.
\newblock {\em Communication of the ACM}, 28:202--208, 1985.

\end{thebibliography}

\end{document}